\documentclass[referee]{aa}

\usepackage{graphicx}
\usepackage{natbib}
\usepackage{chngcntr}
\usepackage{afterpage}
\usepackage{mathtools}
\usepackage{upgreek}
\usepackage{amsmath}

\newcommand{\hs}{\hspace*{0.05cm}}
\newcommand{\norm}[1]{\left\lVert#1\right\rVert}

\begin{document} 


\title{Traces of exomoons in \\
      computed flux and polarization phase curves \\ 
                               of starlight reflected by exoplanets}

\author{J. Berzosa Molina\thanks{Now at GMV AD, Calle de Isaac Newton, 11, 
        28760 Tres Cantos, Madrid, Spain} 
        \and L. Rossi \and D. M. Stam}

\institute{Faculty of Aerospace Engineering, Delft University of Technology,
           Kluyverweg 1, 2629 HS Delft, The Netherlands}

\offprints{J. Berzosa Molina
\email{j.berzosamolina@gmail.com}}

\date{Received 30 April 2018; accepted 23 July, 2018}


\abstract
   {Detecting moons around exoplanets is a major goal
    of current and future observatories. Moons are suspected to influence 
    rocky exoplanet habitability, and gaseous exoplanets in stellar habitable 
    zones could harbor abundant and diverse moons to target in the search for
    extraterrestrial habitats. Exomoons will contribute to exoplanetary 
    signals but are virtually undetectable with current methods.}
   {We identify and analyze traces of exomoons in the temporal variation of
    total and polarized fluxes of starlight reflected by 
    an Earth--like exoplanet and its spatially unresolved moon across all 
    phase angles, with both orbits viewed in an edge--on geometry.} 
   {We compute the total and linearly polarized fluxes, and the
    degree of linear polarization $P$ of starlight that is reflected by 
    the exoplanet
    with its moon along their orbits, accounting for the temporal variation 
    of the visibility of the planetary and lunar disks, and including 
    effects of mutual transits and mutual eclipses.
    Our computations pertain to a wavelength of 450~nm.}
   {Total flux $F$ shows regular dips due to planetary and lunar transits 
    and eclipses.  
    Polarization $P$ shows regular peaks due to planetary transits and 
    lunar eclipses, and $P$ can increase and/or slightly decrease 
    during lunar transits and planetary eclipses. Changes in $F$
    and $P$ will depend on the radii of the planet and moon, on their 
    reflective properties, and their orbits, and are about one magnitude 
    smaller than the smooth background signals. The typical duration of 
    a transit or an eclipse is a few hours.}
   {Traces of an exomoon due to planetary and lunar transits and eclipses
    show up in $F$ and $P$ of sunlight reflected by planet--moon systems
    and could be searched for in exoplanet flux and/or polarization phase 
    functions.}


\titlerunning{Flux and polarization signals of exoplanets with moons} 
\maketitle

\section{Introduction}
\label{sec:introduction}

Since the detection of the first planets beyond 
our Solar System \citep{WolszczanFrail1992,campbell1988search}, 
the number of discoveries has steadily increased, 
yiel\-ding almost 4000 confirmed exoplanets and 2500 unconfirmed, 
candidate exoplanets to this day \citep{han2014exoplanet}. 
Exoplanet space telescopes, such as ESA's CHEOPS 
(CHaracterising ExOPlanet Satellite) and Plato (PLAnetary Transits and 
Oscillations of stars), and NASA's Transiting Exoplanet Survey Satellite (TESS), 
are dedicated to find exoplanets around bright, nearby stars. The relative 
small distances to these stars and their planets combined with the high sensitivity 
of these missions and the upcoming JWST/NASA (James Webb Space Telescope) 
and ARIEL/ESA \citep[][]{2016SPIE.9904E..1XT} missions will allow 
the search for lunar companions and planetary rings.

The continuous increase in instrument precision, stability and spatial 
resolution has enabled a new generation of 
ground--based instruments, such as the Gemini Planet Imager (GPI) 
instrument \citep[see][]{macintosh2014gemini} on the Gemini 
North telescope, the Spectro--Polarimetric High--contrast Exoplanet 
Research (SPHERE) instrument \citep[see][] {beuzit06} on 
ESO's Very Large Telescope (VLT) and the proposed Exoplanet Ima\-ging 
Camera and Spectrograph (EPICS) \citep[see][]{keller10,gratton10} 
on the European Extremely Large Telescope (E-ELT), which is under 
construction by ESO. These high--contrast instruments use 
direct imaging of planetary radiation to not only detect but also
characterize exoplanetary systems through a combination of
spectroscopy and polarimetry techniques. Using GPI and SPHERE, 
respectively, \citet{Macintosh15} and \citet{Wagner673} announced 
the discoveries of young, and thus hot Jovian planets, whose atmospheric 
pro\-perties and orbits were characterized using near--infrared spectroscopy. 

Direct detection of extrasolar bodies presents a major cha\-llenge as 
their observed radiation, both emitted and reflected, is very weak 
compared to that of the host star. In addition, the angular distance 
from the planet to the star is extremely small. Consequently, the 
vast majority of exoplanets have only been detected indirectly. In 
contrast, direct imaging of the planet can reveal a wealth of 
information of the planet properties that cannot be obtained through other 
methods, such as lower atmospheric composition and, for rocky planets, 
their surface coverage.

Despite not having been exploited yet, great hope is placed in 
the polarimetry capabilities of current and future telescopes as 
powerful tools for detecting and characterizing exoplanets 
\citep[see][]{Snik2013, Stametal04, Seageretal00, houghetal03, 
houghlucas03, Saar03, stam03, Stam08}. 
Previous works in this 
field involve the modeling of stellar polarization during 
planetary transits \citep{wiktorowicz2014toward, 
kostogryz2011polarimetric, kostogryz2015polarization,
sengupta2016polarimetric,2005ApJ...635..570C}, 
the modeling of light curves and 
polarization of starlight reflected signals in the visible range 
of Earth-like planets \citep{Rossi2017,Karalidi12,Stam08}, and 
giant Jupiter-like planets 
\citep{Stametal04,Seageretal00}, as well as the mode\-ling of 
exoplanetary atmospheres in the infrared \citep{de2011,MarleySengupta2011}, 
which demonstrated the usefulness of direct observations on exoplanet 
characterization. More recently, \citet{bott2016polarization} reported 
linear polarization observations of the hot Jupiter system HD 
189733, and \citet{2018arXiv180502261G} announced the detection
of planetary thermal radiation that is polarized upon reflection
by circumstellar dust. 
Indeed, polarization has also been proposed as a 
means for exo\-moon detection: \citet{sengupta2016detecting} studied 
the effects of a satellite transiting its hot host planet in the 
polarization signal of (infrared) thermally emitted radiation for 
the case of homogeneous, spherically symme\-tric cloudy planets. 
Studying exomoons 
can improve our understanding of in particular: 

\vspace{-0.2cm}
\begin{enumerate}
\item \textbf{Planet formation:} Solar System moons appear 
to support diverse formation histories. For instance, 
Titan might have formed from circumplanetary debris, 
while the Moon, Phobos and Deimos suggest a cumulative bombardment 
\citep{rufu2017multiple,rosenblatt2016accretion}. 
Triton might have been captured by Neptune
\citep{agnor2006neptune}, while collisions are thought to have 
altered the relative alignment between Uranus and its moons 
\citep{morbidelli2012explaining}. Indeed, studying Solar System 
moons gives essential insights on formation mechanisms and evolution 
\citep[see][and references therein]{heller2017detection}. Exomoon 
research would allow refining planet formation theories 
in a way not achievable by studying exoplanets alone.
\vspace{0.15cm}
\item \textbf{Extra-solar system characterization:} studying 
exomoons will not only provide information on lunar orbits and 
physical properties, but will also allow constraining planet 
characteristics such as i.e. mass, oblateness, 
and rotation axis \citep{barnes2003measuring,Kipping09a,
schneider2015next}. A signal of a planet-moon system could 
be interpreted as that of a planet alone, resulting in e.g.\ an 
overestimation of the planet mass and effective temperature 
\citep{williams2004looking}, and/or an anomalous composition 
from spectroscopy \citep{schneider2015next}. 
Extrasolar system charac\-terization would indeed require 
analysis of all of its elements, i.e.\ planets, 
moons, rings, and exozodiacal dust.  
\vspace{0.15cm}
\item \textbf{Exoplanet and exomoon habitability:}  
a moon may influence its planet's habitability  
\citep{Benn2001}, and moons of giant exoplanets within the stellar 
habitable zone (HZ) might host habitable environments 
\citep{canup2006common}. 
\citet{reynolds1987europa} and \citet{heller2015runaway} mention 
the role of a moon's orbit on the presence of liquid, life--supporting 
water. Indeed, tidal heating could maintain surface temperatures 
compatible with life on large moons around cold giant planets 
\citep{scharf2006potential}. 
\citet{lehmer2017longevity} show that small moons 
could retain atmospheres over limited time periods, 
while Ganymede--sized moons in a stellar HZ could hold  
atmospheres and surface water indefinitely. Although radia\-tion in 
a giant planet's magnetic field and eclipses could threaten 
local conditions for life 
\citep{heller2013exomoon,heller2012exomoon,forgan2014effect},
exomoons are interes\-ting targets in the search for extraterrestrial life.
\end{enumerate}
\vspace{-0.1cm}

\noindent
Led by Kipping's \textit{Hunt for Exomoons with Kepler}, which 
uses a combination of photometric transits, Transit Timing Varia\-tions 
(TTV) and Transit Duration Variations (TDV) data 
\citep{kipping2015hunt,kipping2009transit, sartoretti1999detection, 
szabo2006possibility, simon2007determination, simon2015cheops},
and Hippke's search using the Orbital Sampling Effect (OSE) 
\citep{hippke2015detection,heller2014detecting, heller2016modeling}, 
the search for exomoons is in its starting phase. Mars--sized and 
possibly even Ganymede-sized satellites could be 
traceable in archived Kepler data \citep{heller2014formation}. 
Unfortunately, as of yet no exomoons have been confirmed.
    
In this paper, we use numerical simulations to 
show how an exomoon could influence 
the flux and degree of polarization of the starlight that is reflected 
by an Earth-like exoplanet, using the following outline. 
In Sect.~\ref{sec:rad}, we describe the numerical code to compute 
the various geometries of the exoplanet--exomoon system that are required 
for our radiative transfer computations and the radiative transfer 
code to compute the reflected fluxes and polarization for a given 
exoplanet--exomoon system. 
In Sect.~\ref{sec:results}, we present computed flux and polarization 
phase functions at 450~nm, for an Earth-like planet 
(with a Lambertian reflecting surface and a gaseous atmosphere) with a 
Moon--like satellite (with a Lambertian reflecting surface)
in an edge--on geometry.
Finally, in Sect.~\ref{sec:conclusions}, we summarize and discuss our findings
and their implications.

\section{Computing the reflected starlight}
\label{sec:rad}

\subsection{Stokes vectors and polarization}
\label{sec:stokes}

We describe the flux and polarization of starlight that is reflected 
by a body, with a Stokes vector \citep[see e.g.][]{Hansen1974}:
\begin{equation}
   \mathbf{F} = \left[ {\begin{array}{*{20}{c}}
                    F \\ Q \\ U \\ V
\end{array}} \right],
\label{eq:stokes}
\end{equation}
with $F$ the total flux, $Q$ and $U$ the linearly polarized fluxes, 
and $V$ the circularly polarized flux, all with dimensions W~m$^{-2}$. 
In principle, these fluxes are wavelength dependent. 
However, we will not explicitly include the wavelength in 
the dimensions, because we focus on a single wavelength region.
Fluxes $Q$ and $U$ are defined with respect to a 
reference plane, for which we use the planetary (or lunar) scattering plane, 
which contains the observer, and the centers of the planet (or moon) and the 
star. We do not compute the circularly polarized flux $V$, because 
it is usua\-lly much smaller than the linearly polarized fluxes 
\citep[see][]{RossiStam2018,KAWATA1978217,Hansen1974}, and because
ignoring $V$ does not 
lead to significant errors in the computation of $F$, $Q$, and $U$ 
\citep[see][]{stametal05}. The light of the star is assumed to be 
unpola\-rized \citep[see][]{KempHensonSteinerEtAl1987}, and is 
given by $\mathbf{F}_{\rm 0} = F_0\mathbf{1}$, with $\mathbf{1}$ the 
unit column vector and $\pi F_0$ the flux measured perpendi\-cular 
to the light's propagation direction. If the orbit of the barycenter
of the planet--moon system around the star is eccentric, 
the incident flux varies along the orbit.
Our standard stellar flux, $\pi F_0$, is defined with respect to the 
periapsis of the orbit of the system's barycenter.

The degree of linear polarization, $P$, of vector $\mathbf{F}$ is defined as
\begin{equation}
   P = \sqrt{Q^2+U^2}/F,
\label{eq_pol}
\end{equation}
and the direction of polarization, $\chi$, with respect to the
reference plane can be computed from
\begin{equation}
   \tan{2\chi} = U/Q,
\end{equation}
where $\chi$ is chosen such that $0 \le \chi < \pi$, while 
$\cos{2\chi}$ and $Q$ have the same sign 
\citep{Hansen1974,hovenier2004transfer}.

\subsection{Disk--integrated reflected Stokes vectors}
\label{sec:diskintegration}

We compute the reflected Stokes vector $\mathbf{F}$ of the spatially unresolved 
planet--moon system as a summation of the reflected Stokes vectors 
$\mathbf{F}^{\rm p}$ and $\mathbf{F}^{\rm m}$ of, respectively, the planet 
and the moon (the pair is spatially resolved from the star):
\begin{equation}
   \mathbf{F} = 
       \hs \mathbf{F}^{\rm p} + 
       \frac{R^2_{\rm m}}{R^2_{\rm p}}
       \hs \mathbf{L} (\psi) \hs \mathbf{F}^{\rm m}.
\label{eq:summation}
\end{equation}

Vectors $\mathbf{F}^{\rm p}$ and $\mathbf{F}^{\rm m}$ are disk--integrated
vectors that include the effects of eclipses and transits. 
They are norma\-lized such that the total fluxes reflected by 
the planet and moon at a phase angle $\alpha= 0^\circ$ and without
shadows and/or eclipses on their disks, 
equal the planet's and moon's geometric albedo's, res\-pectively 
\citep[see][]{stam2006integrating}.
Furthermore, $R_{\rm p}$ and $R_{\rm m}$ are the radii of the (spherical) 
planet and moon, respectively.

Vectors $\mathbf{F}$ and $\mathbf{F}^{\rm p}$ in Eq.~\ref{eq:summation}
are defined with 
respect to the planetary scattering plane, while $\mathbf{F}^{\rm m}$ is
defined with respect to the lunar scattering plane. 
Depending on the orientation of the lunar orbit, the lunar scattering plane 
can have a different orientation than the planetary scatte\-ring plane.
Matrix ${\bf L}$ in Eq.~\ref{eq:summation} rotates  
$\mathbf{F}^{\rm m}$ from the lunar to the planetary scattering plane. 
It is given by \citep[see][]{HoveniervanderMee83}
\begin{equation}	
\label{eq:rotation}
   \mathbf{L}(\psi) = \left[ {\begin{array}{*{20}{c}}
                                   1 & 0 & 0 & 0 \\
                                   0 & {\cos{2\psi}} & {\sin{2\psi}} & 0 \\
                                   0 & {\sin{2\psi}} & {\cos{2\psi}} & 0\\
                                   0 & 0 & 0 & 1
                           \end{array}} \right]\,,
\end{equation}
with $\psi$ the rotation angle measured in the clockwise direction from
the lunar to the planetary scattering plane when looking towards the moon 
($0 \le \psi < \pi$).\footnote{\citet{HoveniervanderMee83} define $\psi$  
while rotating in the anti–clockwise direction when looking 
towards the observer, which yields the same angle.}

\begin{figure}[t]
\centering
\includegraphics[width=9cm]{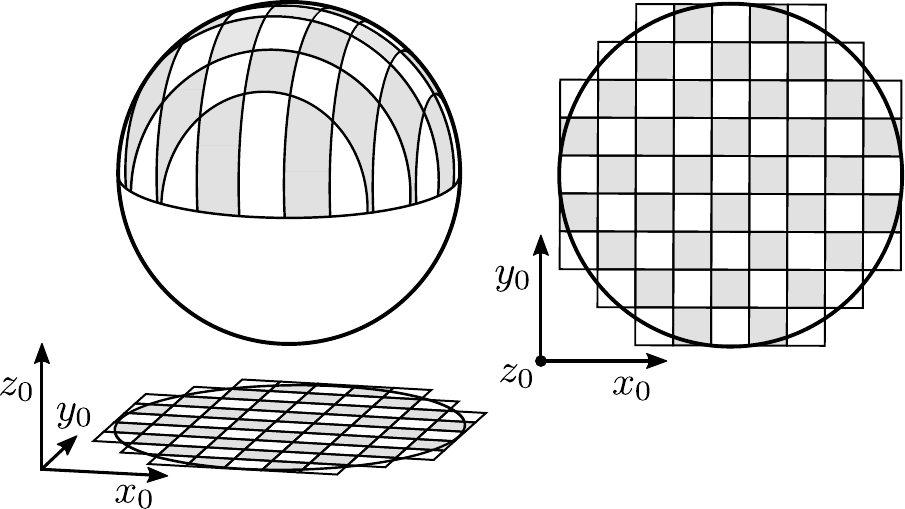}
\caption{3D--view (left) and projection onto the $x_0-y_0$ 
         plane (right) of the discretized planet or moon. The $z_0$--axis points
         towards the observer. The orientation of the $x_0$ and $y_0$ axes with
         respect to the disk of the planet or moon can be chosen arbitrarily.}
\label{fig:grid1}
\end{figure}

To compute the disk--integrated vectors $\mathbf{F}^{\rm p}$ and $\mathbf{F}^{\rm m}$,
we divide the disks of the planet and the 
moon as seen by the observer, into a grid of equally sized, square pixels
(see Fig.~\ref{fig:grid1}).
The number of pixels on the planetary disk is $N_{\rm p}$ 
and that on the lunar disk $N_{\rm m}$. A given pixel will contribute
to a disk--signal when its center is within the disk--radius.
Obviously, the larger the number of pixels (and the smaller each pixel),
the better the approximation of the curved limb of the disk, 
the terminator, and the shadows, such as those due to eclipses (see
App.~\ref{app:NN} for insight into the effect of the number of pixels on 
the computed signals).
The disk--integrated vectors are obtained by summing up the contributions
of the individual pixels across the disk, fully taking into account shadows and 
eclipses, i.e.
\begin{equation} 
{\mathbf{F}}^{\rm x} = 
    \frac{\pi}{N_{\rm x}} \hs
    \displaystyle\sum_{i=1}^{N_{\rm x}} b_i \hs c_i \hs d_i \hs \mathbf{L}(\beta_i) 
    \hs \mathbf{F}^{\rm x}_i,
\label{eq:disk}
\end{equation}
where 'x' is either 'p' or 'm'. Factor $\pi/N_{\rm x}$ is the 
surface area per pixel. 
Furthermore, ${\mathbf{F}}^{\rm x}_i$ is the reflected Stokes vector 
for the $i$-th pixel on the planet (x=p) or moon (x=m), the computation 
of which is described in Sect.~\ref{sec:locallight}.
Matrix $\mathbf{L}$ is a rotation matrix (see Eq.~\ref{eq:rotation}) that is
used for rotating the local Stokes vector ${\mathbf{F}}^{\rm x}_i$
that is defined with respect to the local reference plane, 
to the planetary or lunar scattering plane.
Factor $b_i$ accounts for the visibility of pixel $i$: 
if $b_i = 1$, the pixel is visible to the observer, and if $b_i = 0$
it is invisible due to a transiting body. 
Factor $c_i$ accounts for the dimming of the local incident stellar 
flux due to a (partial) eclipse: $c_i= 0.0$ indicates that pixel $i$ is 
eclipsed and receives no flux, and $c_i=1.0$ indicates that the pixel
is not eclipsed. For partial (penumbral) eclipses, $0.0 < c_i < 1.0$. 
The computation of factors $b_i$ and $c_i$ is
described in {Sect.~\ref{sec:illu}.
Factor $d_i$, finally, indicates the decrease of the standard incident 
stellar flux $\pi F_0$ due to an increase of the distance to the star, according to 
\begin{equation}
d_i = \left( r_{\rm ref} / r_{i{\rm s}} \right)^2,
\end{equation}
where $r_{\rm ref}$ is the reference distance at which the standard 
stellar flux is defined and $r_{i{\rm s}}$ is the actual distance 
between pixel $i$ and the star.

\subsection{The locally reflected starlight}
\label{sec:locallight}

The Stokes vector of the starlight that is reflected by
pixel $i$ on the planet or moon is computed using \citep[see][]{Hansen1974}:
\begin{equation}
   {\mathbf{F}_i^{\rm x}}(\theta_i,\theta_{0i},\phi_i-\phi_{0i}) = 
          \cos{\theta_{0i}}~
          {\bf \mathbf{R}}_{1i}^{\rm x}(\theta_i,\theta_{0i},\phi_i-\phi_{0i})~
          F_0,
\label{eq:refStokes}
\end{equation}
with $\theta_i$ the angle between the local zenith direction and the local 
direction to the observer, $\theta_{0i}$ the angle between the local 
zenith direction and the local direction to the star, and 
$\phi_i - \phi_{0i}$ the local azimuthal difference angle, i.e.\ the angle 
between the plane containing the local zenith direction and the local
direction to the observer and the plane containing the local 
zenith direction and the local direction to the star 
\citep[see][]{rossi2018pymiedap,deHaan87}.  
Furthermore, ${\mathbf{R}}^{\rm x}_{1i}$ is the first 
column of the local reflection matrix of the planet or moon.
Only the first column is needed because the incident starlight 
is assumed to be unpola\-rized (cf.\ Sect.~\ref{sec:stokes}). 
For a given pixel, the illumination and viewing angles, and thus 
$\mathbf{R}^{\rm x}_{1i}$, depend on the position of the planet or moon
with res\-pect to the star and to each other. 
Local reflection matrix $\mathbf{R}^{\rm x}_i$ also depends 
on the local composition and structure of the atmosphere and/or 
surface of the reflecting body.
We compute reflected starlight for an Earth--Moon--like planetary 
system, keeping the reflection models for the Earth and the moon simple 
to avoid introducing too many details that increase computational times while 
not adding insight into the observable signals.

Our model planet has a flat, Lambertian (i.e.\ isotropically and
depolarizing) reflecting surface with an albedo, $a_{\rm surf}$, of 0.3.
The surface is overlaid by an atmosphere that is
assumed to consist of only gas.
We compute the atmospheric optical thickness
at a given wavelength $\lambda$, using a model atmosphere consisting of 32 layers,
with the ambient pressure and temperature according to a mid-latitude
summer profile \cite{1972McClatchey}. The surface pressure is 1.0~bars.
The molecular scattering optical thickness $b^{\rm m}_{\rm sca}$ of an
atmospheric layer at wavelength $\lambda$ is calculated according to
\begin{equation}
   b^{\rm m}_{\rm sca}(\lambda) = \sigma^{\rm m}_{\rm sca}(\lambda) 
                                  \hspace*{0.2cm} N,
\end{equation}
with $\sigma^{\rm m}_{\rm sca}$ the molecular scattering cross--section
(in m$^2$) and $N$ the molecular column number density (in m$^{-2}$)
of the atmospheric layer. The molecular scattering cross--section is 
calculated according to
\begin{equation}
   \sigma^{\rm m}_{\rm sca}(\lambda) = 
        \frac{24 \pi^3}{N_{\rm L}^2}
        \frac{(n(\lambda)^2 - 1)^2}{(n(\lambda)^2 + 2)^2} 
        \frac{6 + 3 \delta(\lambda)}{6 - 7 \delta(\lambda)}
        \frac{1}{\lambda^4},  
\end{equation}
with $N_{\rm L}$, Loschmidt's number at standard pressure and temperature,
$n$, the wavelength dependent refractive index of dry air under
standard pressure and temperature, and $\delta$, the
depolarization factor of the atmospheric gas 
\citep[see][and references therein for the values that have been chosen
for the various parameters]{Stam08}.
To calculate the molecular column number density $N$, we assume 
hydrostatic equilibrium in each atmospheric layer, thus
\begin{equation}
   N = \frac{\delta p}{m g},
\end{equation}
with $\Delta p$ the difference between the pressure at the bottom and
at the top of the atmospheric layer (in Pa), $m$ the average
molecular mass in the layer (in kg), and $g$ the acceleration of 
gravity (in m s$^{-2}$). The atmospheric optical thickness at a 
given wavelength $\lambda$ is calculated by adding the 
values of $b^{\rm m}_{\rm sca}$ for all atmospheric layers
at that wavelength (note that for a model atmosphere containing
only gas, the radiative transfer of incident sunlight only 
depends on the total optical thickness, not on the vertical 
distribution of the optical thickness).
The total atmospheric optical thickness at 450~nm, the wavelength of our
interest, is 0.14. 
At this wavelength, there is no
significant absorption by atmospheric gases in the Earth's atmosphere
\citep[see][for sample spectra]{Stam08}. The single scattering
albedo of the gaseous molecules can thus be assumed to equal 1.0.
And, at this short wavelength, 
the horizontal inhomogeneities of the Earth's surface and the 
contributions of clouds and aerosol to the reflected signal are relatively small
\citep[see][for simulations of the Earth's signal at 440~nm]{Stam08}.
Our model moon has no atmosphere above its flat, Lambertian  
(i.e. isotropic and depolarizing)
reflecting surface with $a_{\rm surf}= 0.1$ \citep{NASAfacts}.

The computation of the local illumination and viewing geometries 
$\theta_i$, $\theta_{0i}$, and $\phi_i-\phi_{0i}$ is described in 
Appendix~\ref{sec:computing-angles}.
Given these angles and the planet's atmosphere--surface model, 
we use PyMieDAP\footnote{PyMieDAP is freely available under the GNU GPL 
license at {\tt http://gitlab.com/loic.cg.rossi/pymiedap}} 
\citep{rossi2018pymiedap}, an efficient radiative transfer code based 
on the adding--doubling algorithm described by \citet{deHaan87}. 
PyMieDAP fully includes polarization for all orders of scattering,
and assumes a locally plane--parallel atmosphere--surface model
to compute ${\mathbf{R}}_{1i}^{\rm p}$ for every pixel on the planet.
The computed locally reflected Stokes vector, ${\mathbf{F}_{i}}^{\rm p}$, 
is defined with respect to the local meridian plane, 
i.e.\ the plane through the local zenith and the local direction towards the 
observer. For each illuminated pixel on the moon, 
${\mathbf{R}}_{1i}^{\rm m} = a_{\rm surf} \mathbf{1}$.
A detailed description of PyMieDAP including benchmark results
can be found in \citet{rossi2018pymiedap}. 

Results of our radiative transfer code have been compared against results 
presented in e.g.\ 
\citet{Stam08,Stam2006,Stametal04} (who all used the same adding--doubling code, 
but an entirely different disk--integration algorithm), and
\citet{Karalidi12} (who used their own version of an adding--doubling code
and an independent disk--integration method).
\citet{buenzli2009grid} and \citet{2017Stolker}, 
each compared their own,
independently implemented Monte Carlo radiative transfer codes
successfully against results from the code used by 
\citet{Stametal04} and \citet{Stam2006}.

\begin{figure}[b]
\centering
\includegraphics[width=\textwidth]{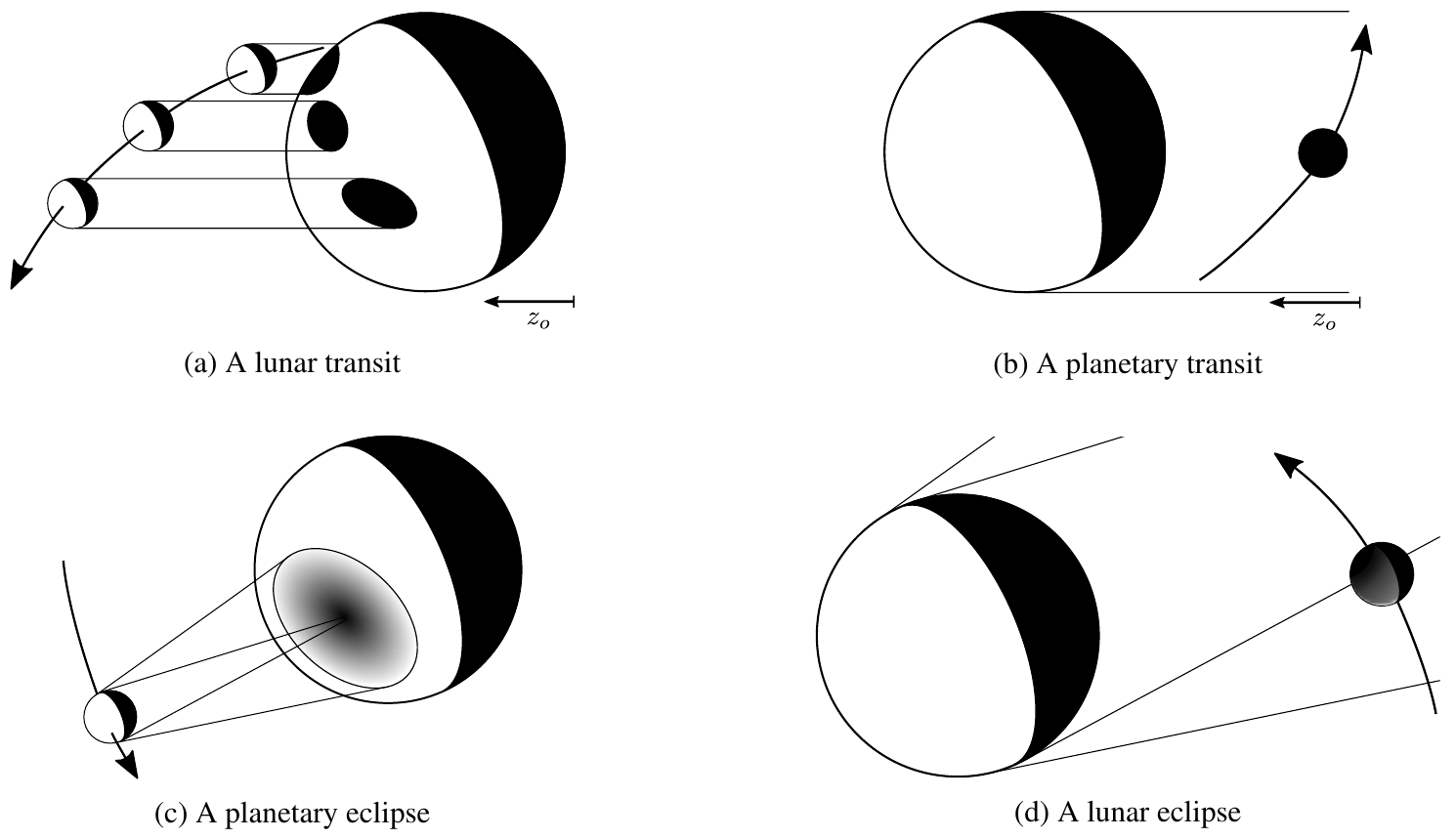}
\caption{Sketch illustrating the mutual events between a planet and its moon:
         (a) a lunar transit, (b) a planetary transit,
         (c) a planetary eclipse, and (d) a lunar eclipse.
         The positive $z_0$--axis (in sub-figures a and b) points to the observer.
         The white--black scale indicates full--null illumination of a body.
         An arrow indicates the lunar motion around the planet.
         The distances and radii are not to scale.
         }
\label{fig:transit-eclipse}
\end{figure}

\subsection{Computing transits and eclipses}
\label{sec:illu}

As described in Eq.~\ref{eq:disk}, the contribution of the light reflected 
by a pixel~$i$ on the planet or the moon to the disk--integrated 
Stokes vector $\mathbf{F}$, 
depends on the factors $b_i$ and $c_i$, that account for the pixel's 
visibility and dimming, respectively. The values of these factors depend on 
so--called mutual events, specifically, transits, in
which one body is (partially) blocking the light that is reflected
towards the observer by another body, and eclipses, in which one
body is casting a (partial) shadow on the illuminated and visible disk
of another body.
Limiting ourselves to systems in which a single star
is orbited by a planet with a single moon, we distinguish the following 
four mutual events (cf.\ Fig.~\ref{fig:transit-eclipse}):
\begin{enumerate}
\item \textbf{A planetary eclipse}: the moon is between the star 
and the planet, casting its shadow on the planet, the extent of 
which depends on the planet--star and moon--star 
distances, on the stellar, planetary and lunar radii, and on their orbital
positions.
\item \textbf{A lunar eclipse}: the planet is between the star and 
the moon, casting its shadow on the moon, the extent of 
which depends on the planet--star and moon--star 
distances, on the stellar, planetary and lunar radii, and on their orbital positions.
\item \textbf{A planetary transit}: the planet is between the 
moon and the observer, blocking the view of the moon, 
the extent of which depends on the planetary and lunar radii, and
their orbital positions. 
\item \textbf{A lunar transit}: the moon is between the planet 
and the observer, occulting a region of the planetary disk,
the extent of which depends on the planetary and lunar radii,
and their orbital positions.
\end{enumerate}

\noindent
We exclude planetary and lunar transits of the star, i.e.\
the epochs in which these bodies move in front or behind the star. 
Numerical simulations of transiting planets with moons 
have been \-published by \citet{Kipping2011}. 
Modeling the transmission and scattering of starlight in the 
planetary atmosphere
during those epochs (which is not included in the work by \citet{Kipping2011}),
requires a fully spherical atmosphere model instead of a locally
plane-parallel one \citep{2012Icar..221..517D} and falls outside
the scope of this paper.

For our computation of the effects of transits of the planet in front
of the moon and vice versa on the flux and polarization of the reflected
starlight, we assume that the bodies are at 'infinite' distance of 
the observer.
For our computation of the effects of the eclipses, i.e. the shadow
of one body darkening regions on the other body, on the reflected flux
and polarization, we fo\-llow the mathematical description of 
eclipses in the Moon--Earth system\- as developed by \citet{link69},
taking into account the sizes of the planet and the moon, their
distances and positions with respect to the star, and the size of the
stellar disk. The latter is crucial for the modeling of the umbra,
antumbra en penumbra shadow regions (for an example of the umbra and
penumbra, see Fig.~\ref{fig:transit-eclipse}). The contribution
of the starlight reflected by pixels in the antumbral or penumbral region
of the planet or moon to the total signal is weighted by the depth
of the shadow (i.e.\ factor $c_i$ in Eq.~\ref{eq:disk}). 
We ignore stellar limb darkening and the transmission of 
starlight through the planetary atmosphere during a lunar eclipse.

A detailed description of our numerical computation of eclipses 
and the factor $c_i$ in Eq.~\ref{eq:disk} 
is provided in Appendix~\ref{sec:eclipses}.
This computation requires the positions of the planet and the moon
with respect to the star across time, and thus the dynamics of the 
three--body system. The basics of this dynamics is outlined in the 
next section.

\subsection{Computing the orbits of the planet \& moon}
\label{sec:orbits}

\begin{figure}[b]
\centering
\includegraphics[width=1.1\textwidth]{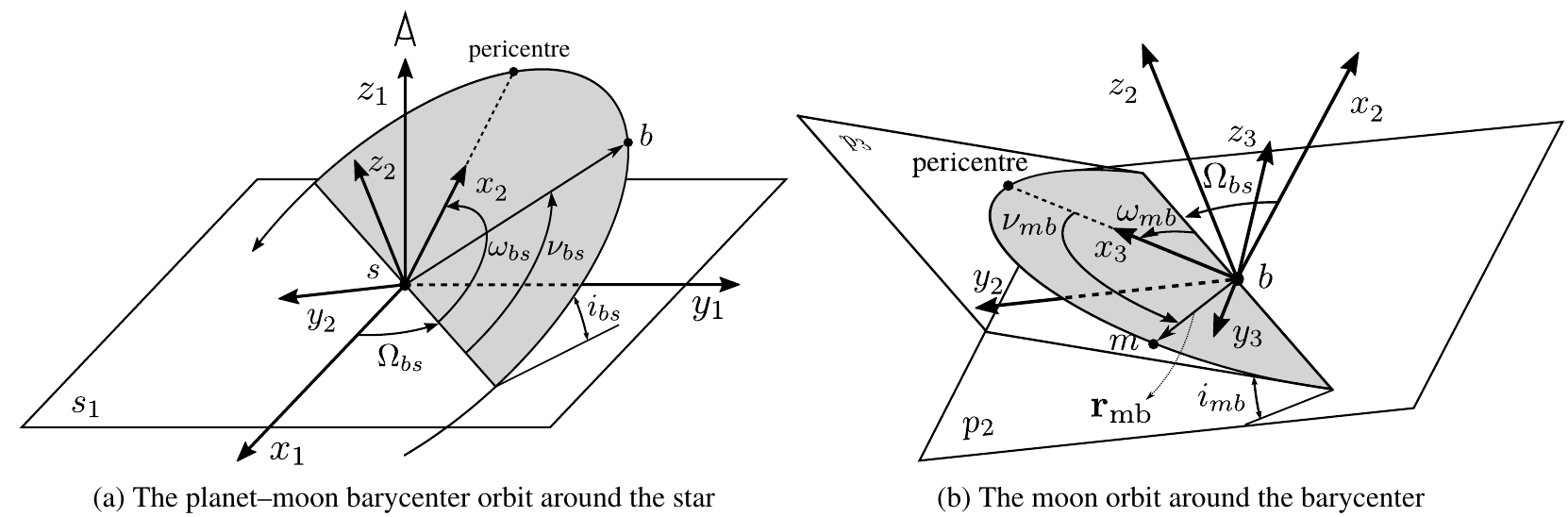}
\caption{Sketch of the reference frames and angles used to describe
         the Keplerian orbits of the
         planet--moon system barycenter around the parent star (Fig.~a) and
         the Keplerian orbit of the moon around the planet--moon system
         barycenter (Fig.~b). Plane $p_1$ (a) is the plane of the sky as seen
         by the observer on the positive $z_1$--axis.
         Plane $p_2$ (Figs.~a \& b) is the barycenter's orbital plane,
         and plane $p_3$ (Fig.~b) is the orbital plane of
         the moon around the barycenter. Angle $i$ is the orbital
         inclination angle, $\omega$ the argument of periastron,
         $\Omega$ the right ascension of the ascending node,
         and $\nu$ the true anomaly. Subscript ${\rm bs}$ refers to the
         barycenter of the planet--moon system around the star,
         and ${\rm mb}$ to the moon around the barycenter.
         Vectors $\mathbf{r}_{\rm bs}$ and $\mathbf{r}_{\rm mb}$ are
         the position vectors of the barycenter and the moon, respectively.
         The barycenter and the moon are indicated by
         $b$ and $m$, respectively.}
\label{fig:ref_frames}
\end{figure}

We compute the position vectors of the planet and moon as functions of time for 
determining the factors $b_i$, $c_i$, and $d_i$ of each pixel $i$
and for evalua\-ting the disk--integration according to Eq.~\ref{eq:disk}.
Both the motions of the planet and its moon around the star
depend on their mutual gravitational interactions. 
Assuming each body attracts as a point mass and neglecting the gravity of other
planets and/or moons in the system, our star--planet--moon 
system is a classical, generic three--body problem. 

A precise computation of the orbital positions in the generic 
three--body problem 
requires the numerical propagation of a given set of initial conditions. 
Instead, we use the 'nested two--body' appro\-ximation
described by \citet{Kipping2011,kipping2010weigh}, which assumes 
that the orbits of the planet and moon around the
planet--moon system barycenter, and the orbit of this 
barycenter around the star can all be described by Keplerian orbits.
The advantages of the nested two--body approximation are:
1.~the solution can be described analytically;
2.~the computational time is significantly shorter than with 
numerical integrations;
3.~it provides better insight in the computed orbits as the
elements of all orbits can be specified;
4.~unlike the circular restricted three--body problem 
simplification \citep[see e.g.]{wakker2015fundamentals}, 
it can handle eccentric orbits.

As demonstrated by \citet{kipping2010weigh}, the nested two--body
approximation is excellent for the generic three--body problem 
provided $\Re \le 0.531$, where $\Re$ is the 
moon--planet separation in units of the planet's Hill's sphere radius
\citep[see e.g.][]{de2015planetary}. 
As follows from \citet{Domingos2006} and \citet{Kipping2011}, 
stable, prograde orbiting moons should \-fulfill 
$\Re \le 0.4895$, while retrograde orbiting moons 
can be stable up to $\Re \approx 0.9309$. 
The nested two--body approximation can thus be applied to all 
prograde orbiting moons, while retrograde orbiting moons are only 
partially \-covered, depending on $\Re$. 
We will limit ourselves to prograde orbiting moons, as we do not 
expect any influence of the moon's orbital direction on the 
magnitude of reflected flux and polarization features, except 
on their timing.

Figure~\ref{fig:ref_frames} shows the geometry of the planet--moon 
system with the reference frames describing the orbit of the planet--moon 
barycenter around the star, and the orbit of the moon around the 
barycenter. The nested two--body approximation assumes that the motions of the
planet--moon barycenter around the star and that of the moon around the 
barycenter are independent.
Orthonormal, right--handed coordinate system 
$S_1=\{x_1,y_1,z_1\}$ is the reference frame for the observation of 
the planet--moon--star system, with the star at the origin, 
and the \emph{$z_1$}--axis pointing towards the observer. 
Plane $p_1$, through $x_1$ and $y_1$, is the plane on the observer's
sky, onto which the pixels (Fig.~\ref{fig:grid1}) are projected. 
Axes \emph{$x_1$} and \emph{$y_1$} have an arbitrary 
(but fixed) orientation. Coordinate system $S_2=\{x_2,y_2,z_2\}$ is the
reference frame for the orbit of the barycenter, which lies in plane $p_2$,
through $x_2$ and $y_2$.
The lunar orbit itself lies in the $x_3$--$y_3$--plane
of coordinate system $S_3=\{x_3,y_3,z_3\}$ that is centered at the 
barycenter's position. 
The various orbital elements in these coordinate systems
are indicated as follows:
\[
\begin{array}{lp{0.8\linewidth}}
         a      & semi--major axis         \\
         e      & eccentricity [0, 1]            \\
         i      & inclination angle [0\degr, 180\degr]             \\
         \omega & argument of periastron [0\degr, 360\degr]   \\
         \Omega & right ascension of the ascending node [0\degr, 360\degr] \\
         \nu    & true anomaly [0\degr, 360\degr]
\end{array}
\]
With the orbital parameters of the barycenter and the moon,  
we compute the true anomalies $\nu_{\rm bs}$ of the barycenter
and $\nu_{\rm mb}$ of the moon around the barycenter, 
respectively, at any time $t$ using Kepler's equation for elliptical orbits,
i.e.\
\begin{equation}
\label{eq:kepler}
   E(t) - e \sin{E(t)} = M(t),
\end{equation}	
where $E$ is the eccentric anomaly and $M$ the mean anomaly of the
orbit.
The true anomalies of the barycenter and the moon are related to
the eccentric anomaly $E$ through:
\begin{equation}
\label{eq:nu-E}
   \tan{\dfrac{\nu(t)}{2}} = \sqrt{\dfrac{1+e}{1-e}}\tan{\dfrac{E(t)}{2}}.
\end{equation}	

We solve for $\nu(t)$ for each orbit in the appropriate reference system
by applying the Newton--Raphson method \citep{wakker2015fundamentals}
to Eqs.~\ref{eq:kepler} and~\ref{eq:nu-E}.

We compute the position of
the barycenter, $\mathbf{r}_{\rm bs}$, in coordinate system $S_2$,
and the position of the moon around the barycenter, 
$\mathbf{r}_{\rm mb}$, in coordinate system $S_3$.
The absolute position of the moon in coordinate system $S_2$
is then obtained through:
\begin{equation}
	\mathbf{r}_{\rm ms}(t) =  \mathbf{r}_{\rm mb}(t) + 
    \mathbf{r}_{\rm bs}(t)\,.
\label{eq:rms-comp}
\end{equation}

As formulated by \citet{Murray10}, the position of the barycenter and
the moon in $S_2$ at time $t$ 
can be put through a series of transformation matrices
to yield the position of the barycenter and the moon in the observer's 
coor\-dinate system $S_1$. 
For further details on these transformation matrices, see  
\citet{Kipping2011,kipping2010weigh}. 

After having computed the positions of the planet and the moon 
in $S_1$ at time $t$, we compute the positions of the pixels\- across the 
planetary and lunar disks (see Fig.~\ref{fig:grid1}), and the angles
$\beta_i$ that are used to rotate locally computed Stokes vectors to 
the planetary and lunar scattering planes (Eq.~\ref{eq:disk}),
respectively,
Then we calculate parameter $d_i$, which accounts for the change
of the standard incident flux due to the changing distance to the star
(see Eq.~\ref{eq:disk}),
and the local illumination and viewing angles required for 
the computation of the Stokes vector of reflected starlight for each 
pixel seen by the observer. Details on these computations can be found in 
App.~\ref{sec:computing-angles}.
For each $t$, we also compute angle $\psi$ to rotate ${\bf F}^{\rm m}$,
the disk-integrated Stokes vector for the moon, to the planetary
scattering plane (Eq.~\ref{eq:summation}).

\subsection{Our baseline planet--moon system}
\label{sec:baseline}

In this paper, we focus on planet--moon systems in edge--on geometries,
in which the inclination angle of the barycenter's orbit is 90$^\circ$,
because exoplanets in (near) edge--on orbits are prime targets for   
space telescopes such as TESS, JWST, PLATO and CHEOPS, that all will employ the 
transit method to detect and/or characterize exoplanets, as well as 
for follow--up missions including telescopes aimed at directly 
detecting planet signals.

\begin{figure}[b]
\centering
\includegraphics[width=0.5\textwidth]{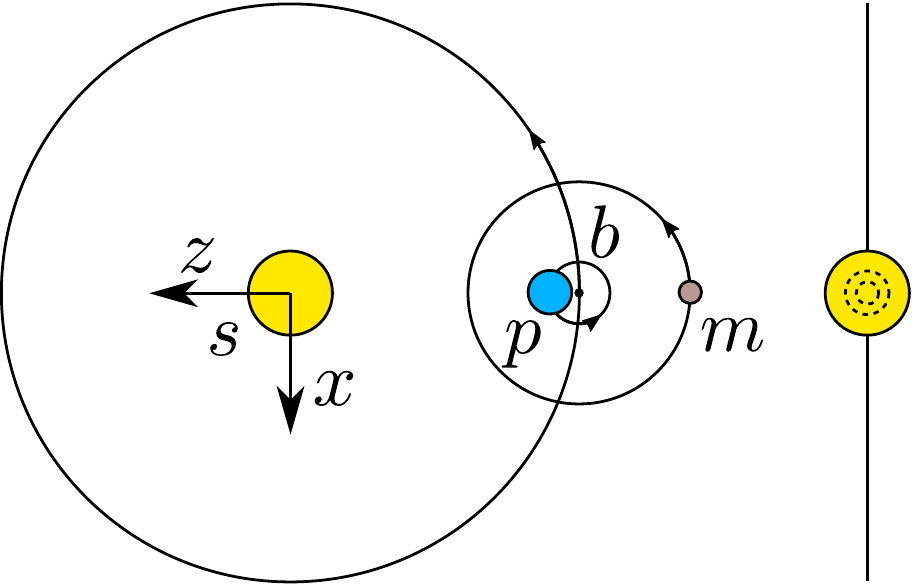}
\caption{Sketch illustrating the orbital geometry of our edge--on
         planet--moon system at time $t=0$ as seen from the positive
         $y$--axis (left) and from the observer's position at the positive
         $z$--axis (right). Indexes $s$, $p$, $m$, and $b$ refer to the
         positions of the star, the planet, the moon, and the planet--moon
         system barycenter, respectively.
         Distances and radii are not to scale.}
\label{fig:orbit_edge_on}
\end{figure}

Table~\ref{tab:edge-on} lists the orbital elements of our baseline 
planet--moon system.
Both the barycenter's and the lunar orbit are assumed\- to be circular ($e=0.0$), 
and their semi--major axes match those of the Earth--Moon system \citep{NASAfacts}. 
We neglect the Earth--barycenter distance, so that $a_{\rm bs}= 1$~AU. 
The semi--major axis of the lunar orbit, $a_{\rm mb}$, is computed from the 
Moon--Earth semi--major axis, $a_{\rm mp} = 2.5696 \cdot 10^{-3}$~AU \citep{NASAfacts}, 
as follows: 
\begin{equation}       
	a_{\rm mb} = a_{\rm mp} \hs \frac{m_{\rm p}}{m_{\rm p} + m_{\rm m}}
                 \approx 2.54 \cdot 10^{-3} \hspace*{0.2cm} {\rm AU} \,,
\label{eq:ampamb}    
\end{equation}
with $m_{\rm p}$ and $m_{\rm m}$ the masses of the Earth and Moon, 
respectively.
Because of the edge--on geometry, the right ascensions of the 
ascending nodes of the orbits of the barycenter and the moon are 
set to zero.
Because both orbits are assumed to be circular, their perihelions 
are undefined. The barycenter's argument of perihelion is chosen 
precisely behind the star at time $t = t_0 = 0$, 
i.e. $\omega_{\rm b} = 270\degr$. 
For the moon, $\omega_{\rm m}$ is set to zero.
The observational and orbital geometry at $t=0$ is sketched in 
Fig.~\ref{fig:orbit_edge_on}.
We use $R_{\rm p}= 6371.0$~km and $R_{\rm m}= 1737.4$~km for
the baseline radii of the planet and the moon, respectively.

\begin{table}[t!]
\centering
\begin{tabular}{l|cc}
 ~          		& Barycenter  & Moon     \\ \hline
$a$  ~[AU]   		&       1.0   &    0.00254     \\
$e$  ~[-]    		&   	0.0 	  &    0.0     \\
$i$~~~[\degr]    		&   	90.0 	  &    0.0     \\
$\omega$ [\degr]    	&   	270.0 	  &    0.0     \\
$\Omega$ [\degr]    	&   	0.0 	  &    0.0     \\
$t_0$ [sec] 	    	&   	0.0 	  &    0.0    
\end{tabular}
\caption{The orbital elements of the barycenter of our planet--moon system
         and of the orbit of the moon, with $a$ the semi--major axis, 
         $e$ the eccentricity, $i$ the inclination angle, $\omega$ the 
         argument of the perihelion, $\Omega$ the right ascension of the 
         ascending node, and $t_0$ the time of perihelion passage.
         The inclination angle of the lunar orbit is defined with 
         respect to the normal on the barycenter orbit.}
\label{tab:edge-on}
\end{table}

\begin{figure}[tbh]
\centering
\includegraphics[width=1\textwidth]{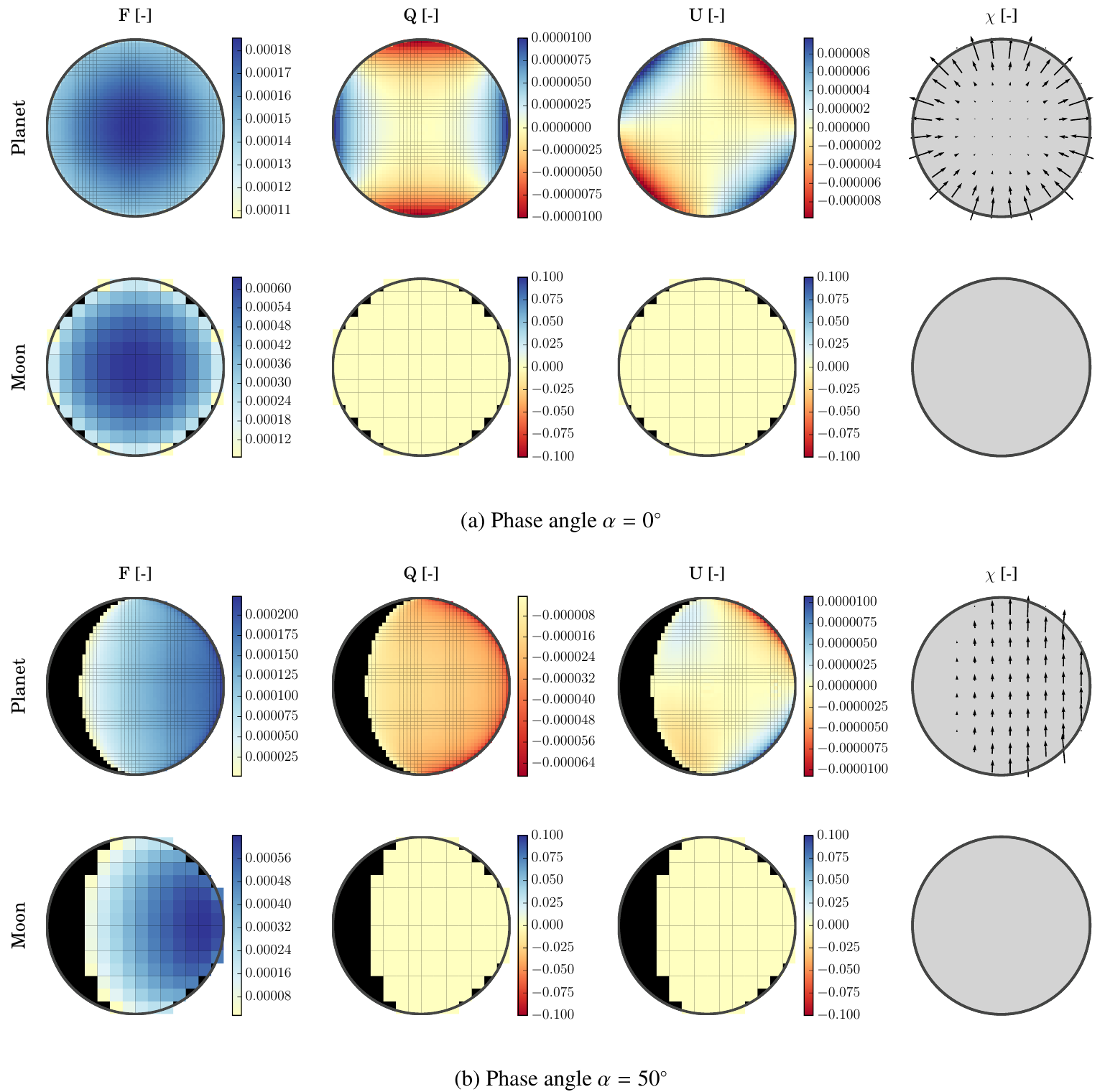}
\caption{Reflected fluxes $F$, $Q$, and $U$, and the direction of polarization
         $\chi$ across the planet and the moon at $\alpha=0\degr$ (top) 
         and~50$\degr$ (bottom). Fluxes $Q$ and $U$, and $\chi$
         are defined with respect to the scattering planes of the planet and 
         the moon, respectively. In order to facilitate the interpretation of the
         degree of polarization, we plot $180\degr-\chi$. Fluxes $Q$ and $U$ 
         are zero across the moon's disk because of the Lambertian reflection.
         All fluxes have been normalized such that the disk--integrated 
         flux $F$ at $\alpha= 0\degr$ equals the body's geometric albedo. 
         Absolute planetary fluxes per pixel are not comparable to the
         absolute lunar fluxes per pixel
         because of the different number of pixels across each disk.}
\label{fig:IQUPPixel}
\end{figure}

\section{Numerical results}
\label{sec:results}

Here, we present the computed total flux $F$, the linearly polarized fluxes
$Q$ and $U$, and degree of polarization $P$ of starlight that is reflected 
by our model planet--moon system across time. 
As a trade--off between spatial resolution, radiometric and polarimetric 
accuracy, and computational time, we use 50 and 14 pixels along the equators
of the planet and moon, respectively (see App.~\ref{app:NN}), resulting in
$N_{\rm p}=1956$ and $N_{\rm m}=156$ (Eq.~\ref{eq:disk}).
In Sect.~\ref{sec:r1-ref_prop}, we analyze the individual contributions of the 
planet and the moon, and in Sect.~\ref{sec:r1-system}, the results for the 
spatially unresolved planet--moon system. In Sect.~\ref{sec:tran},
we take a closer look at particular transit and eclipse events.

\subsection{Reflection by the spatially resolved planet {\&} moon}
\label{sec:r1-ref_prop}

In order to understand the traces of eclipses and transits in the flux and
polarization of starlight reflected by spatially unresolved planet--moon systems, 
we first discuss the disk--resolved signals of the 
planet and the moon separately. Figure~\ref{fig:IQUPPixel} 
shows the elements of the locally reflected Stokes vectors 
$\mathbf{F}^{\rm p}$ and $\mathbf{F}^{\rm m}$ and the direction 
of polarization $\chi$, with respect to the planetary and lunar 
scattering planes, respectively,   
at phase angles $\alpha$ of 0$\degr$ and 50$\degr$.

At $\alpha=0\degr$ (Fig.~\ref{fig:IQUPPixel}a), 
both the planet and the moon would be behind the star
and thus invisible, but their disk--resolved signals give 
insight in the reflection processes. For both bodies, total flux $F$ is 
maximum at the sub-stellar/sub-observer region
and decreases towards the terminator (which coincides with the limb at
this phase angle). Because of the Lambertian reflection of 
the lunar surface and the lack of atmosphere around the moon, the reflected
flux is unpolarized and $\chi$ undefined for the moon. The linearly polarized 
fluxes $Q$ and $U$ of the planet are due to Rayleigh scattering in the 
planet's atmosphere. At the sub-stellar region, both $Q$ and $U$ are zero
because of symmetry. The general increase of $Q$ and $U$ towards the limb 
is due to polarized second order scattered light, which is also apparent from the 
direction of polarization $\chi$. Because of its definition, $Q$ ($U$)
equals zero along the lines at angles of 45$^\circ$ (0$^\circ$) 
and -45$^\circ$ (90$^\circ$) with the horizontal. Integrated across the
planetary disk, $P$ would equal zero. 
Note that because of the Lambertian reflection 
of the surface of the planet, $Q$ and $U$ are independent of planetary surface 
albedo $a_{\rm surf}$, 
while $P$ will generally decrease with increasing $a_{\rm surf}$ because of 
the increasing flux $F$ \citep[see e.g.][for sample computations]{Stam08}.

At $\alpha= 50^\circ$ (Fig.~\ref{fig:IQUPPixel}b), 
the total flux $F$ of the moon is maximum at 
the sub-stellar region and decreases towards the terminator, 
due to the isotropic surface reflection and the absence of an atmosphere. 
The planet also shows a decrease of $F$ towards the terminator,
but the location of the flux maximum is more diffuse and more towards
the limb than on the moon, because light that is incident on the planet
is scattered in the atmosphere in addition to being reflected by the surface; 
the reflected flux thus also depends on the
optical path--lengths through the atmosphere, which in turn depend on the local 
illumination and viewing angles. 
The planet's polarized fluxes $Q$ and $U$, and angle $\chi$ are mostly 
determined by starlight that has been singly scattered by the atmospheric
gas molecules. For our choice of reference plane, $U$ is anti--symmetric
with respect to this plane (and $U$ would thus equal zero when 
integrated across the
disk), and $Q$ is symmetric. The negative values for $Q$ in 
Fig.~\ref{fig:IQUPPixel}b indicate that the reflected light is polarized 
perpendicular to the reference plane, which is
indeed also clear from the polarization angle $\chi$,
and what is expected for a Rayleigh scattering atmosphere.

\begin{figure*}[t]
\centering
\includegraphics[width=0.85\textwidth]{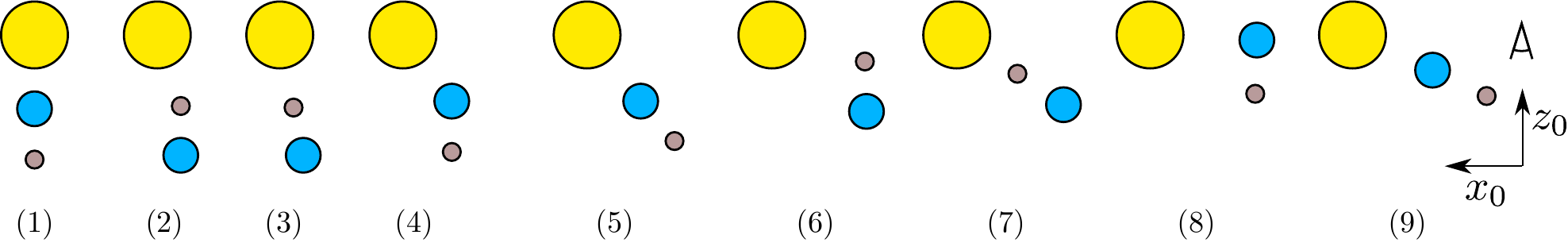}
\caption{Sketch illustrating the sequence of planetary (1, 4, 8, ...) and
         lunar (2, 6, ...) transits, as well as planetary (3, 7, ...) and
         lunar (1, 5, 9, ...) eclipses for part of the barycenter's
         orbit for an edge-on system. The positive $z_0$--axis points
         towards the observer. Position 1 corresponds to phase angle
         $\alpha=0^\circ$ and time $t=0$~s in our simulation
         (cf.\ Fig.~\ref{fig:ref_edge_on}).}
\label{fig:edge_on_series}
\end{figure*}

\begin{figure*}[p]
\centering
\includegraphics[width=1\textwidth]{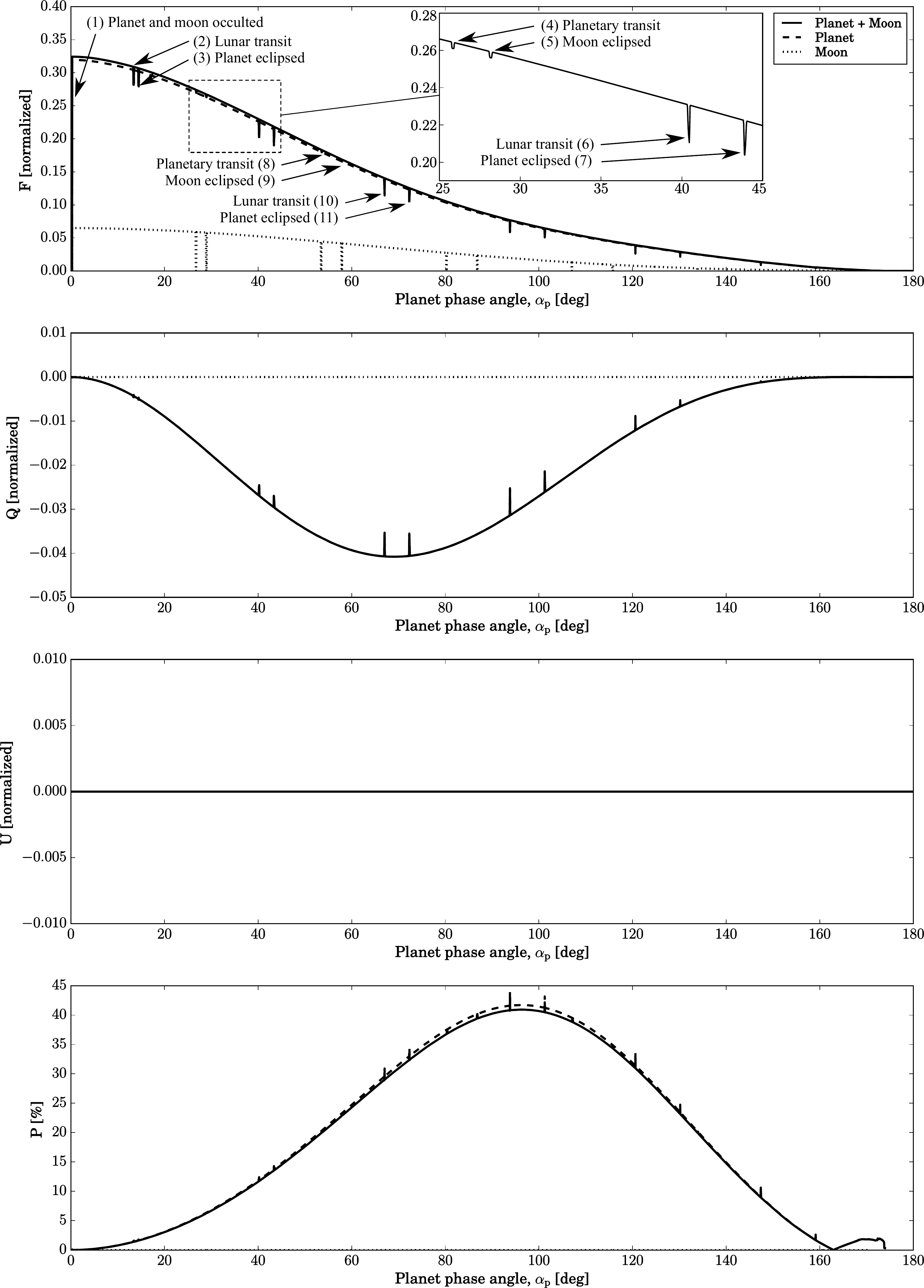}
\caption{The total flux $F$, the linearly polarized fluxes $Q$ and $U$,
         and the degree of polarization $P$ of the spatially unresolved,
         edge--on, base-line planet--moon system, as functions of the
         planet's phase angle
         $\alpha_{\rm p}$. Also included are curves for the isolated planet
         (coinciding with $Q$ and $U$ of the planet--moon system)
         and the isolated moon (equal to zero in $Q$, $U$, and $P$).
         Fluxes have been normalized such that at $\alpha_{\rm p}= 0\degr$,
         $F$ equals the geometric albedo of the planet--moon system
         or each of the isolated bodies.
         The labels in the plot for $F$ refer to the illustrations
         in Fig.~\ref{fig:edge_on_series}.}
\label{fig:ref_edge_on}
\end{figure*}
\subsection{Reflection by the spatially unresolved planet {\&} moon}
\label{sec:r1-system}

Figure~\ref{fig:ref_edge_on} shows the disk--integrated $F$, $Q$, $U$, and $P$
as functions of the planetary phase angle $\alpha_{\rm p}$.  
For our edge--on system, the curves cover only half of the barycenter's 
orbital period.
For comparison, we have also included curves for the planet and the
moon as isolated bodies, thus without any mutual events.
The total flux of the planet--moon system is lower than the sum
of the total fluxes of the isolated planet and moon because the latter
have not been scaled to the actual radii of the planet and the moon.
Indeed the moon's flux at $\alpha_{\rm p}=0^\circ$, equals the moon's geometric
albedo, i.e.\ 0.067, which matches the theoretical geometric albedo 
of a Lambertian reflecting body with surface albedo of 0.1
\citep[see][]{Stam2006}. The geometric albedo of the unresolved 
planet--moon system (with both bodies at $\alpha_{\rm p}=0^\circ$ and next to
each other) is about 0.33. 
Note that in Fig.~\ref{fig:ref_edge_on}, the {\em observable} 
planet--moon flux at $\alpha_{\rm p}=0^\circ$ is zero, because both bodies
are then located behind the star as seen from the observer 
(in addition, the moon is located behind the planet,
as can be seen from Fig.~\ref{fig:orbit_edge_on}, and from situation~1
in Fig.~\ref{fig:edge_on_series}).

The curves for $F$ decrease smoothly with increasing $\alpha_{\rm p}$, apart
from the occasional sharp dips due to eclipses and transits (to be discussed 
below) and reach zero close to $\alpha_{\rm p}= 180^\circ$, where the planet
and moon would both be in front of the star. The slightly different slope of 
the lunar flux phase function as compared to that of the planet is due to 
the scattering of light in the latter's atmosphere. 
As can be seen in Fig.~\ref{fig:ref_edge_on}, without the sharp dips, the smooth
flux phase function of the planet--moon system does not reveal the presence 
of a moon, especially not without accurate information on
the planetary radius, orbital distance, atmospheric and surface properties. 

Because the lunar surface is completely depolarizing, 
the moon's polarized fluxes $Q$ and $U$ are zero at each 
$\alpha_{\rm p}$. The disk--integrated $U$ of the light 
reflected by the planet is zero due to symmetry (see Fig.~\ref{fig:IQUPPixel}).
Polarized flux $Q$ and degree of polarization $P$ of this light both 
show a smooth dependence on $\alpha_{\rm p}$, apart from the occasional 
sharp peaks that will be discussed below.
The degree of polarization is maximum at phase angles between 90$^\circ$
and 100$^\circ$ due to the atmospheric Rayleigh scattering.
At about 165$^\circ$, the direction of polarization changes from perpendicular 
($\chi=90^\circ$) to parallel ($\chi=0^\circ$) to the reference plane, 
and $P$ equals zero. 
The degree of polarization of the unresolved planet--moon system 
is somewhat lower than that of the isolated planet, because of the 
added unpolarized lunar flux. 
If the planetary atmosphere would contain clouds, 
the shape of this continuum curve would depend on the optical thickness 
and altitude of the clouds, the microphysical properties of the particles, and the
cloud coverage across the planetary disk 
\citep[for sample curves, see][and references therein]{Rossi2017,Karalidi12}.

While the smooth curves for the spatially unresolved planet--moon system
shown in Fig.~\ref{fig:ref_edge_on}, do not provide direct evidence of the presence 
of a moon, the mutual events result in a series of dips and peaks in the 
reflected flux and polarization, respectively. 
Figure~\ref{fig:edge_on_series} illustrates the various events. 
Both the planet and the moon are initially ($\alpha_{\rm p}=0^\circ$) 
behind the star (position 1 in Fig.~\ref{fig:edge_on_series}).
Given the prograde lunar motion, the next event, when planet and moon
are in view of the observer, is a lunar transit (position 2) and an eclipse of the 
star on the planet (3). After the first lunar period, the moon again 
disappears behind the planet (4), followed by an eclipse of the star on the 
moon (5). This sequence repeats along the barycenter's orbit.

Both the planetary and lunar eclipses and transits temporarily reduce the 
flux $F$ that the observer receives. Indeed, when the planet transits
the moon, the system's flux phase function equals that of the isolated 
planet. 
The dip in the system's flux due to a lunar transit 
(moon in front of the planet) will depend 
on the radius of the moon as compared to that of the planet 
and on the lunar surface albedo: the lower the lunar 
surface albedo and/or the larger the lunar radius, the deeper the dip 
compared to the continuum. 
The depth of the dip in the system's flux $F$ due to an eclipse
depends on the relative sizes of the moon and the planet, the reflection  
properties of the eclipsed body, and on the 
precise orbital geometry, especially because an eclipse shadow on the moon will not 
be completely black (cf.\ Fig.~\ref{fig:transit-eclipse}) (and the total flux $F$
thus slightly larger) due to starlight that is refracted 
through the limb of the planetary atmosphere and reaches the moon. This refraction is not 
included in our code (due to the wavelength dependence of Rayleigh scattering, 
the contribution of refracted light would be larger in the (near) infrared 
region of the spectrum than at 450 nm).

Because the moon reflects unpolarized light, 
neither a planetary transit (planet in front of the moon) nor an 
eclipse on the moon leads to a reduction
of the polarized fluxes, as can be seen in Fig.~\ref{fig:ref_edge_on}.
Because the planet reflects polarized light, a transit
of the moon and an eclipse on the planet will both decrease $Q$
(given the geometry of our system).
Because $P$ depends on $F$ and $Q$, the dips in $F$ due to less 
(unpolarized) lunar light being observed yield peaks in $P$. 
The peak value of $P$ that 
is due to the planet transiting the moon equals $P$ of the isolated planet
at that value of $\alpha_{\rm p}$. 
In our computational results, 
peaks in $P$ that are due to an eclipse on the moon,
would equal $P$ of the isolated planet when the whole lunar disk would be 
in the planet's umbra because we neglect refracted starlight through the 
limb of the planetary atmosphere.
Changes in $P$ that are due to the moon transiting the
planet or due to the moon casting a shadow on the planet will depend on the
total and polarized fluxes of the region of the planetary disk 
that is covered or darkened, and thus, for a given model planet and its
atmosphere, on the 
relative sizes of the moon and the planet and the precise orbital geometry. 
This will be discussed further in Sect.~\ref{sec:tran}.

The absolute depth of the dips in $F$ and $Q$ decreases 
with increasing $\alpha_{\rm p}$ because the fraction of a body's disk that 
is illuminated and visible decreases with increasing $\alpha_{\rm p}$.
The amplitudes of features in $P$ for our planet-moon model system are maximum when
$\alpha_{\rm p} \approx 90\degr$. This is particularly convenient for exomoon 
detection with direct imaging techniques, because that is the phase
angle range where the angular distance between the planet--moon system
and the parent star will be largest. 

As can be seen in Fig.~\ref{fig:ref_edge_on}, the phase angle gap 
between a lunar transit and the subsequent planetary eclipse (or a 
planetary transit and a subsequent lunar eclipse) increases with increasing 
$\alpha_{\rm p}$. As the orbital speed of both bodies is constant in our
baseline system with circular orbits, this also applies in the time domain. 
Indeed, the lunar and planetary transits have a characteristic
period because an observer--planet--moon alignment occurs 
twice per lunar orbit (see Fig.~\ref{fig:edge_on_series}). 
The time gap between two consecutive transits and eclipses, however, 
increases with increasing $\alpha_{\rm p}$ because of the movement 
of the barycenter along its orbit.

\subsection{Analysis of the mutual events}
\label{sec:tran}

In this section, we analyze individual mutual events, i.e.\ their shape, 
symmetry, periodicity, magnitude, and duration.  
For this analysis, the change in flux $F$ and degree
of polarization $P$ during an event are defined as follows
\begin{eqnarray}
   \Delta F & = & F_{\rm event} - F_{\rm continuum}, \\
   \Delta P & = &  P_{\rm event} - P_{\rm continuum}.
\end{eqnarray}
First, we'll discuss the lunar and planetary transits, then the 
lunar and planetary eclipses. 
\\

\begin{figure}[b]
\centering
\includegraphics[width=0.5\textwidth]{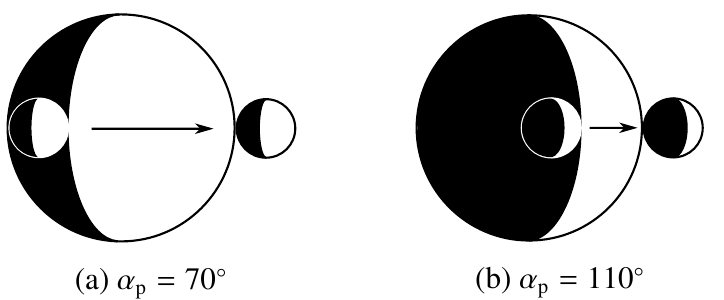}
\caption{Sketch of the ingress and egress of the moon during a lunar transit
         for $\alpha_{\rm p} = 70\degr$ and $\alpha_{\rm p}=110\degr$. 
         The arrow indicates 
         the direction of motion of the moon across the planetary disk.}
\label{fig:tt}
\end{figure}

\noindent {\bf Lunar transits}
\vspace*{0.2cm}

\noindent 
Figures~\ref{fig:lt_both}a sand b provide detailed views of $\Delta F$ and 
$\Delta P$ during the six lunar transits (moon in front of the planet) 
shown in Fig.~\ref{fig:ref_edge_on},
together with sketches of the geometry of the planet and the moon at
the beginning and the end of the transit for $\alpha_{\rm p} \approx 80^\circ$.
As expected with constant orbital speeds, 
the duration of a lunar transit event decreases with increasing $\alpha_{\rm p}$ 
because of the decrease of the illuminated area on the planetary disk, 
and thus the shift of the time of ingress (see Fig.~\ref{fig:tt}). 
Because egress takes place over the planetary limb, all curves in 
Figs.~\ref{fig:lt_both}a and b have the same egress time. 
Also, the planet is relatively dark near the terminator, and thus 
yields a smooth flux decrease upon the lunar ingress, while it is bright
near the limb (see Fig.~\ref{fig:IQUPPixel}), yielding a rapid increase of 
$F$ upon the lunar egress. 

The depth $\Delta F$ depends strongly on $\alpha_{\rm p}$, 
because with increasing $\alpha_{\rm p}$, the illuminated area, and 
hence also the covered area on the planet decreases. 
The shape of $\Delta F$ also depends on $\alpha_{\rm p}$.
At $\alpha_{\rm p}=0^\circ$, the curve would be symmetric.
At larger values of $\alpha_{\rm p}$, the trace of the 
lunar night--side starts to appear in the curve. 
Because of the moon's prograde orbit, the lunar day--side ingresses before 
the night--side. In the curves for $\alpha_{\rm p}=13.4^\circ$,
and 40.3$^\circ$, the steeper decrease of $F$ due to the ingress of the
lunar night--side can be seen. The value of $\Delta F$ that is
reached within a transit at a given $\alpha_{\rm p}$ depends on the lunar 
albedo and on the area of the planetary disk that is covered, thus on
the lunar radius. 
The lower the lunar surface albedo and/or the
larger the lunar radius as compared to the planetary radius, 
the larger $\Delta F$ will be. 
As an example, Fig.~\ref{fig:tR}a shows $F$ at 
$\alpha_{\rm p}= 67.2^\circ$ for various values of the lunar radius 
expressed as fraction of the planetary radius (the value for the
baseline model is approximately 0.3). 
As can be seen, the continuum flux increases with increasing lunar radius
due to the increased amount of flux reflected by the moon, and, indeed the 
lowest flux during the transit decreases and $\Delta F$ increases
with increasing lunar radius.

Note that a change in the lunar radius implies a change in the lunar mass 
(assuming a similar composition) and, thus, a change in the lunar period 
around the planet. While the frequency of the events decreases non--linearly 
with increasing lunar radii, we have aligned the mutual events in 
Fig.~\ref{fig:tR} in time to facilitate a comparison. 
Mutual transits show up every half lunar sidereal period. 
Because $T \propto \sqrt{1/(M_{\rm m}+M_{\rm p})}$, relative timing variations 
of 1\% to 13\% are obtained for lunar--to--planet radius ratios from 
0.1 to 0.7 (with our baseline value of approximately 0.3). 
In the case of eclipses, and assuming coplanar circular lunar and 
planetary orbits, the repetition period equals the lunar synodic period, 
for which timing variations of 1\% to 14\% are obtained for the same 
range of radii ratios.

\begin{figure}[t]
\centering
\includegraphics[width=1\textwidth]{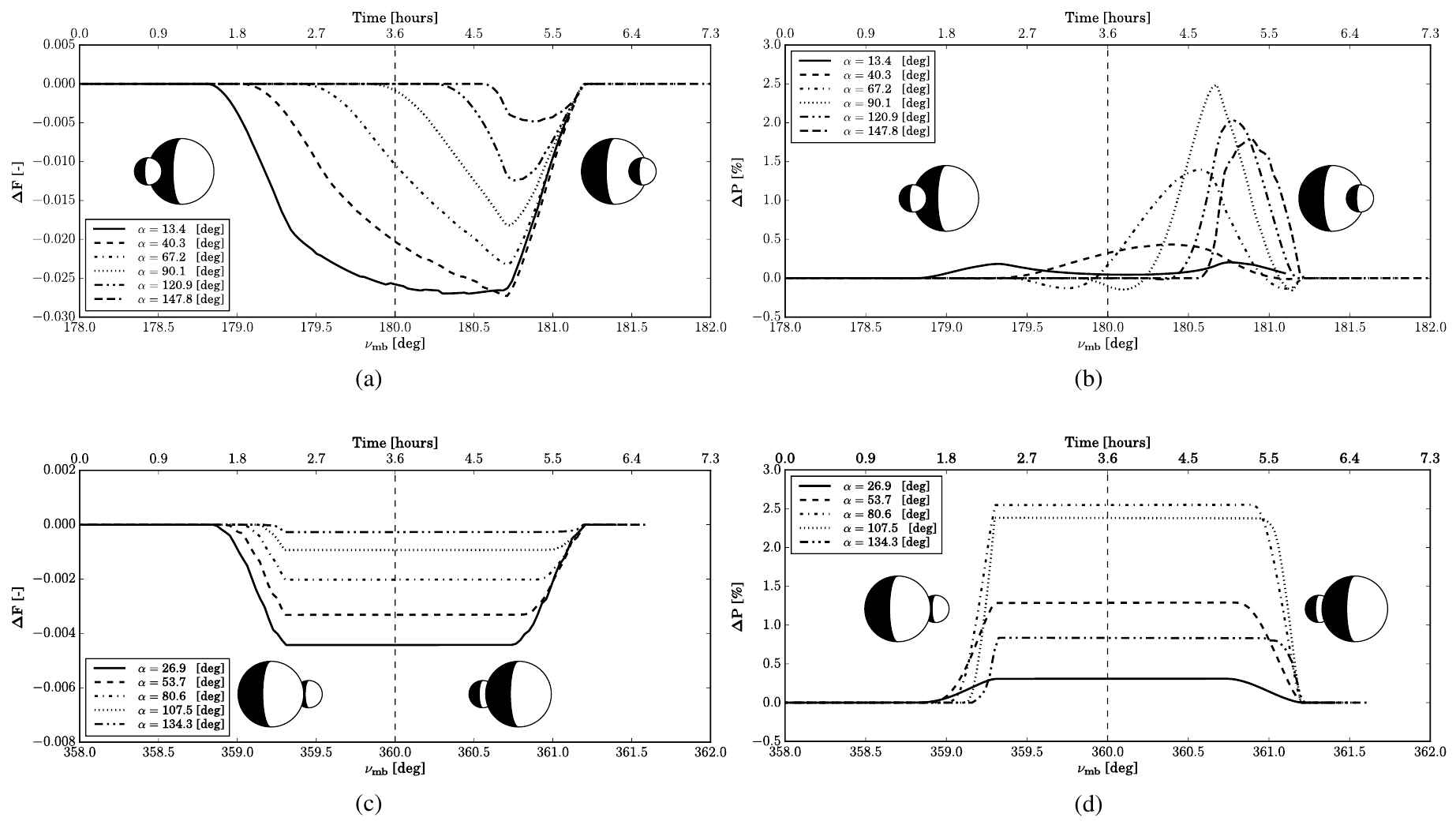}
\caption{Changes in the total reflected flux $\Delta F$ (a and c), and the 
         degree of polarization $\Delta P$ (b and d), as functions of the 
         lunar true anomaly, $\nu_{\rm mb}$, and relative time 
         for the lunar transits (top) and planetary transits (bottom) 
         shown in Fig.~\ref{fig:ref_edge_on}.
         The time--step of these simulations is 3 minutes.}
\label{fig:lt_both}
\end{figure}
\begin{figure}[t]
\centering
\includegraphics[width=1\textwidth]{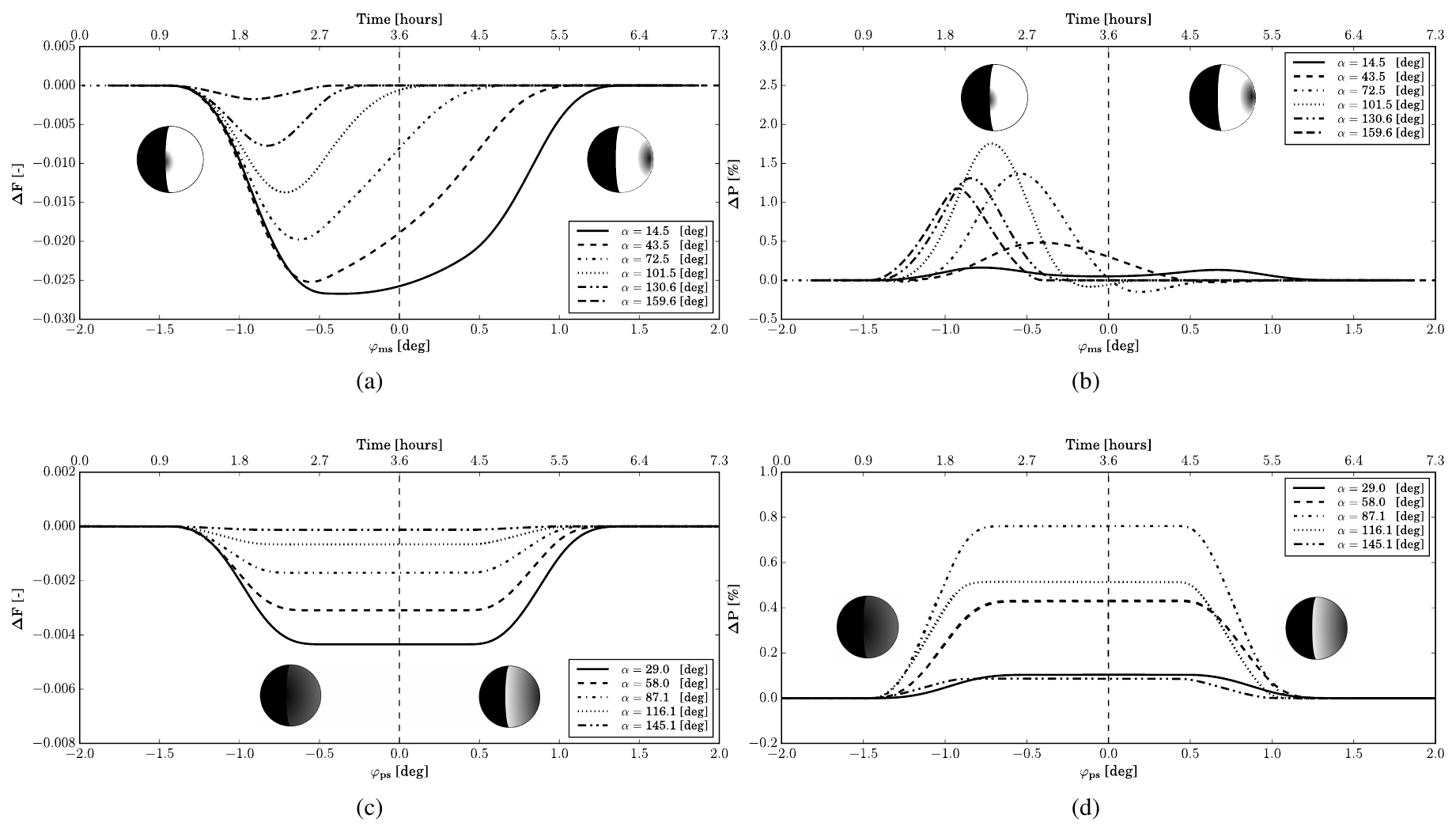}
\caption{Similar to Fig.~\ref{fig:lt_both}, except for the planetary
             eclipses (top) and the lunar eclipses (bottom), both as functions of  
             angle $\varphi_{\rm ms}$ (see Fig.~\ref{fig:eclipsevarphi}).}
\label{fig:le_both}
\end{figure}
\begin{figure}[t]
\centering
\includegraphics[width=1\textwidth]{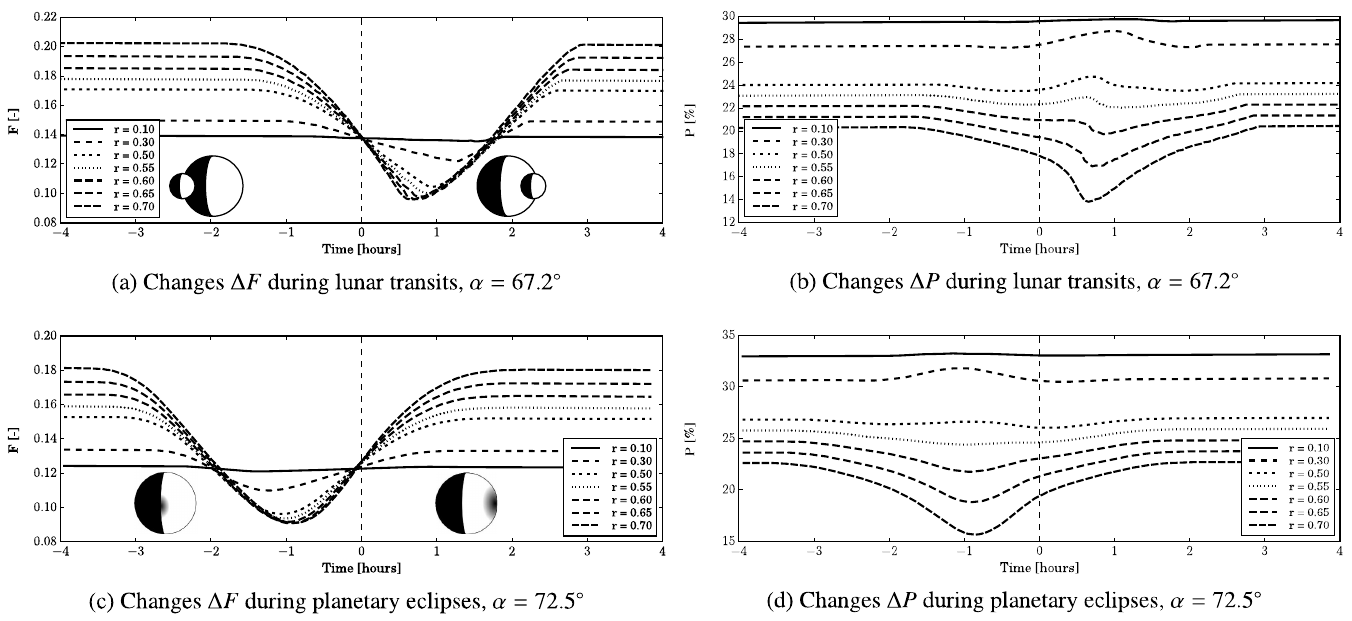}
\caption{Changes in the total reflected flux $\Delta F$ (a and c) and $\Delta P$ (b and d)
         during lunar transits at $\alpha = 67.2\degr$ (top) and planetary eclipses
         at $\alpha = 72.5\degr$ (bottom) for various lunar--to--planetary radius
         ratios $r$. The horizontal axis shows the elapsed time since the
         concentric alignment of the planet and moon as seen from the star
         in the case of an eclipse and as seen from the observer in the case of a
         transit. The time--step of these simulations is 3 minutes.
         The baseline lunar--to--planetary radius ratio $r$ is about 0.3.}
\label{fig:tR}
\end{figure}

Figure~\ref{fig:lt_both}b shows $\Delta P$ during lunar transits.
It can be seen that $P$ can also decrease during a transit, which is 
not apparent from the curves in Fig.~\ref{fig:ref_edge_on}.
The curves exhibit a strong variation in shapes, and with increasing 
$\alpha_{\rm p}$, get increasingly asymmetric. 
The largest $\Delta P$ is found around $\alpha_{\rm p}=90^\circ$, 
where $P$ of our model planet is highest
(see Fig.~\ref{fig:ref_edge_on}).
The precise shapes of the curves depend on the properties of the 
planet and its moon and the path of the transit across the planetary disk.

In our planet--moon system, the lunar transit occurs along the planet's 
equator, where the antisymmetry of $U$ yields a null net contribution, 
and the shape of $\Delta P$ thus depends on the variation of $Q$ and $F$
along the path (cf.\ Fig.~\ref{fig:IQUPPixel}).
At $\alpha_{\rm p}=13.4^\circ$, $Q$ is maximum near the planet's limb and 
close to zero at the center of the body. The disk--integrated $P$ is close to 
zero due to symmetry. 
During ingress and egress, the moon breaks the symmetry, 
and (slightly) increases the disk--integrated value of $P$ (see Fig.~\ref{fig:lt_both}b). 
With increasing $\alpha_{\rm p}$, the maximum $\Delta P$ increases, 
to reach a maximum (in the figure) at $\alpha_{\rm p}=90.1^\circ$.
The $\alpha_{\rm p}$ where the maximum $P$ is found, corresponds roughly with
the $\alpha_{\rm p}$ of the minimum $F$. This increase in $P$ appears to be 
driven by the decrease of $F$.

The negative values of $\Delta P$ in the curves for $\alpha_{\rm p}=67.2^\circ$,
$90.1^\circ$ and 120.9$^\circ$, indicate that during that part of the transit, 
the decrease in $Q$ is larger than that in $F$. 
This happens in particular when the illuminated part of the moon is transiting 
the illuminated part of the planetary disk, while the dark part of the moon
is still transiting the dark part of the planet, and when the moon transits the
limb of the planet, where $Q$ is relatively large and where the transit thus
strongly decreases $Q$.  

As mentioned earlier, the maximum $\Delta P$ (on the order of 2~\% 
for our planet--moon system) should be observable when $\alpha_{\rm p}$ is 
between about 70$^\circ$ and 150$^\circ$, where the angular distance 
between the planet--moon and the star is relatively
large, which would facilitate the detection of lunar transits with 
direct detection methods. 

Figure~\ref{fig:tR}b shows the change in $P$ at
$\alpha_{\rm p}=67.2^\circ$, for various values of $r$, the lunar radius
expressed as a fraction of the planetary radius.
With increasing $r$, the continuum $P$ decreases because more
unpolarized flux reflected by the moon is added to the total flux.
During a transit, $P$ can be seen to be very sensitive to the lunar 
radius. Indeed, with increasing $r$, $\Delta P$ changes from positive 
($P$ during the event is higher than in the continuum) to negative 
($P$ during the event is lower than in the
continuum) because apparently, the polarized flux reflected by
the planet decreases more during the event than the total flux reflected 
by the planet with the moon in front of it.
\\

\noindent {\bf Planetary transits} 
\vspace*{0.2cm}

\noindent Figures~\ref{fig:lt_both}c and d show the planetary transits 
(planet in front of the moon) from Fig.~\ref{fig:ref_edge_on}
in detail, both for $F$ and $P$.
At $\alpha_{\rm p}=0^\circ$, the planet and the moon are behind the star 
and would thus not be visible to the observer, but, similarly to the lunar 
transits, 
planetary transits would yield symmetric events in $\Delta F$ and $\Delta P$.
With increasing $\alpha_{\rm p}$, the events become more asymmetric. 

With a planetary transit, the shapes of the $\Delta F$ curves are very 
different from those of the lunar transits, because, for our orbital geometry,
the moon will be completely covered during the planetary transit and while it is
covered, the transit curve is flat. The depth of the transit depends
on the total amount of reflected flux received
from the (isolated) moon. 
For a given value of $\alpha_{\rm p}$, $\Delta F$ will increase 
linearly with the surface area of the lunar disk (thus, with the lunar radius
squared), and/or with the lunar surface albedo.
The start time of the transit depends on $\alpha_{\rm p}$, as 
$\alpha_{\rm p}$ determines the extent of the illuminated area on the moon,
and thus when it will be covered. 
Like with the lunar transits, the end of the transit, over the bright
limb of the moon, is independent of $\alpha_{\rm p}$; it only depends on 
the lunar true anomaly.

For the planetary transits, $\Delta P$
is entirely due to the decrease in $F$, as the moon itself
reflects only unpolarized light, and the planet thus blocks no polarized flux. 
As a result, the transits in $\Delta P$ are flat as long as the illuminated 
part of the lunar disk is covered.
The maximum $\Delta P$ depends both on the $\Delta F$ and on the planet's
polarized flux $Q$, and thus on $\alpha_{\rm p}$ for a given planet-moon
model.  
In Fig.~\ref{fig:lt_both}d, a maximum $\Delta P$ of about 
2.5$~\%$ occurs at $\alpha_{\rm p}=80.6\degr$. 
Through $\Delta F$, $\Delta P$ will increase 
with the lunar surface albedo and/or the surface area of the lunar disk
at a given $\alpha_{\rm p}$ and for a given planet-moon model; a darker 
and/or smaller moon would yield a smaller $\Delta F$ and hence a smaller 
$\Delta P$. 
\\

\noindent{\bf Planetary eclipses} 
\vspace{0.2cm}

\noindent Figures~\ref{fig:le_both}a and b show the curves of $\Delta F$ 
and $\Delta P$ during the planetary eclipse events shown before in 
Fig.~\ref{fig:ref_edge_on}. During these eclipses, the moon casts its
shadow on the planet, and because in our geometry the lunar orbital plane
coincides with the barycenter's orbital plane, the shadow of the moon 
travels along the horizontal line crossing the center of the 
planetary disk.
Figure~\ref{fig:eclipsevarphi} illustrates the geometries for the 
planetary eclipse, with $\varphi_{\rm ms}$ the angle between the star 
and the moon measured positive in the counter clock--wise direction
from the center of the planet. Angle $\varphi_{\rm ms}$ is used 
as relative measure of eclipse events. Its relation with time is
linear, given the circular orbital motion of the bodies.

For planetary eclipses, the explanation regarding the asymmetry of 
$\Delta F$ and $\Delta P$ is the same as for lunar transits, with one 
important difference: while the transit events start increasingly 
later with increasing $\alpha_{\rm p}$ and end at the same (relative) time, 
the planetary eclipses start at the same (relative) time and end 
increasingly earlier with increasing $\alpha_{\rm p}$.
The reason for this difference is that eclipses depend on the position
of the star with respect to the planet--moon system, and not on the 
position of the observer. The observer's position does influence the 
fraction of the eclipse that is captured, as it determines the
phase angle and hence the fraction of the illuminated part of the planetary
disk across which the eclipse travels.
Thus, with increasing $\alpha_{\rm p}$, the duration of an eclipse
decreases, as is visible in Figs.~\ref{fig:le_both}a and b.
The depth $\Delta F$ decreases with increasing $\alpha_{\rm p}$ because less
of the illuminated part of the planetary disk is visible.

The shape of the $\Delta F$ curves for the planetary eclipses (where the 
moon's shadow moves across the planetary disk) appears to be more 
gradual than that of the lunar transit curves (where the moon itself
moves across the planetary disk) (cf.\ Fig.~\ref{fig:lt_both}). 
This is because the planet first travels through the lunar penumbral shadow, 
before entering the deep, umbral shadow cone.
Because we discuss only half of the barycenter's orbit, the ingress of the lunar 
eclipse shadow on the illuminated part of the planetary disk is
through the terminator and the egress through the bright limb, as illustrated 
in Figs.~\ref{fig:tt} and~\ref{fig:eclipsevarphi}. However, because the 
spatial extent of the penumbral and umbral shadows across the disk are 
smallest halfway the total 
duration of the eclipse (as seen from a vantage point on the moon facing the 
planet), the egress of the lunar shadow yields a much smoother $\Delta F$
curve than observed during the egress of a lunar transit. 
The difference in the maximum $\Delta F$ during a lunar transit and a planetary 
eclipse is most apparent at the smaller phase angles, because with increasing 
$\alpha_{\rm p}$, the contribution of light reflected by the moon decreases.
Note that differently than for lunar transits, the value of $\Delta F$ during a
planetary eclipse
is independent of the lunar surface albedo. It will obviously increase
with the radius of the moon relative to that of the planet. 
This can also be seen in Figs.~\ref{fig:tR}c and d,
which show $F$ at $\alpha_{\rm p}=72.5^\circ$ for various values
of $r$, the lunar radius expressed as fraction of the planetary radius
(for the baseline model, $r$ is 0.3). Indeed, with increasing $r$, 
the continuum flux increases because of the added flux reflected by the 
moon, and the minimum flux during the event decreases because of the 
increasing extent of the lunar shadow.

The change in $P$ during the planetary eclipses is shown in Fig.~\ref{fig:le_both}b.
The increase in $P$ during the ingress of the lunar shadow is due to
the decrease in $F$ and a decrease in $Q$. 
As the shadow progresses across the disk, its spatial
extent decreases, and its influence on $P$ decreases. 
The maximum of $\Delta P$ appears to be $1 - 2~\%$ 
for phase angles $\alpha_{\rm p} \approx 70\degr-160\degr$.
For the largest phase angles, the corresponding value of $\Delta F$ 
is relatively small, because only a narrow crescent of the planet
is illuminated, so there $\Delta P$ is mostly due to a 
change in $Q$. 

Figure~\ref{fig:tR}d shows $P$ at 
$\alpha_{\rm p}=72.5^\circ$ for various values of $r$ (the lunar radius
expressed as fraction of the planetary radius). With increasing $r$, 
the continuum $P$ decreases because of the added unpolarized flux reflected 
by the moon, and $P$ during the eclipse decreases, too, apparently because the
polarized flux $Q$ decreases more than the total flux $F$.
\\

\begin{figure}[b]
\centering
\includegraphics[width=0.52\textwidth]{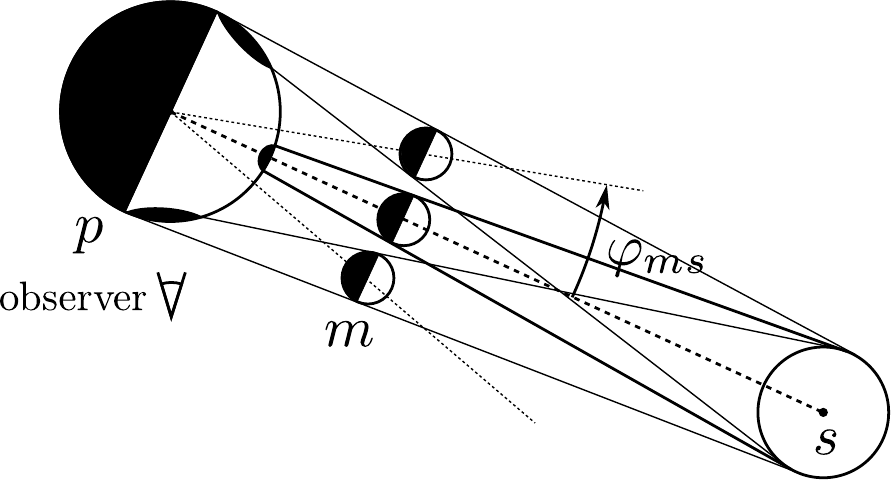}
\caption{The geometrical definition of angle $\varphi_{\rm ms}$ as moon $m$
         passes between planet $p$ and star $s$. Seen from the top,
         the moon moves anti--clockwise around the planet. A similar definition
         holds for angle $\varphi_{\rm mp}$ as the planet passes between the 
         moon and the star. 
         Distances between bodies and radii are not to scale.}
\label{fig:eclipsevarphi}
\end{figure}

\noindent {\bf Lunar eclipses} 
\vspace{0.2cm}

\noindent As can be seen in Figs.~\ref{fig:le_both}c and d, lunar eclipses, 
when the planet casts its shadow on the moon, show a similar symmetry 
as the planetary transits, where the planet moves in front of the moon
(Fig.~\ref{fig:lt_both}c and d). Because our model moon is small compared to the 
planetary shadow, both the $\Delta F$ and $\Delta P$ curves
are flat except during ingress and egress. The flux changes
during ingress and egress of the lunar eclipse, 
respectively, are smoother than with the planetary transits, 
due to the extended penumbral region of the planet's shadow.

Like with the planetary eclipses discussed above, the moon's ingress 
into the planetary shadow occurs at the same value of 
angle $\varphi_{\rm ms}$ (cf. Fig.~\ref{fig:eclipsevarphi}) independent
of phase angle $\alpha_{\rm p}$. The duration of the eclipse decreases with 
increasing $\alpha_{\rm p}$ because of the decrease of the illuminated area 
on the moon with increasing $\alpha_{\rm p}$.
The change in $P$ during lunar eclipses, shown in Fig.~\ref{fig:le_both}, 
is similar to that during planetary transits (Fig.~\ref{fig:lt_both}), 
as in both cases $P$ changes during the event because $F$ decreases
and because there is 
no actual change in the amount of polarized flux from the system, 
because our model moon reflects only unpolarized light. 
The maximum $\Delta P$ value is about $2.7~\%$, 
attained at $\alpha_{\rm p} \approx 87.1^\circ$ in Fig.~\ref{fig:le_both}.

With increasing lunar radius as compared to the planetary 
radius, and depending on the distance between the moon and the planet 
and the distance to the star, the planetary shadow might cover only
part of the lunar disk. In that case, $F$ and $P$ will no longer be
flat during the eclipse, because there will be a
contribution of unpolarized lunar flux that will vary in time.
Indeed, the curves will become asymmetric (more similar to those for the 
planetary eclipses). Because the moon only reflects unpolarized
flux, $P$ will always increase during the eclipses.

\section{Conclusions}
\label{sec:conclusions}

We present numerically simulated flux and polarization phase functions of 
starlight that is reflected by an orbiting planet--moon system, including 
mutual events, such as transits and eclipses. 
Most results presented in this paper apply to a Moon--sized, 
Lambertian (i.e. isotropically and depolarizing) reflecting moon 
orbiting an Earth--sized exoplanet with an Earth--like, 
gaseous atmosphere on top of a Lambertian reflecting
surface (the surface pressure is 1 bar), 
in an edge--on configuration. 
Our results show that the flux and polarization
phase functions of starlight reflected by such a planet--moon 
system contain traces of the moon in the form of periodic changes in the 
total flux $F$ and degree of polarization $P$ as the bodies shadow each 
other (eclipses) and/or hide one another from the observer's view (transits) 
along their orbit around the star. 
These changes in $F$ and $P$ are only one order 
of magnitude smaller than the system's continuum phase functions.
The magnitude, shape and duration of the 
obtained total flux signatures are comparable with the results by 
\citet{cabrera2007detecting}, except that they do not include 
the influence of the penumbra. 

During events that darken the planet, i.e.\ the lunar transits and planetary 
eclipses, the shape of the dip in $F$ depends on the 
reflection properties of the regions on the planet along the path of the shadow. 
The change in $P$ during such events strongly depends 
on the ratio of polarized--to--total reflected flux across the disk
and along the path of the shadow. Indeed, $\Delta P \approx 1\% - 1.8 \%$ for 
$67\degr < \alpha_{\rm p} < 121\degr$. For the planet--moon system used in 
this paper, 
we found the strongest changes in $P$ during either the first (planetary eclipse)
or the last (lunar transit) half of the event, compared to the duration of
the event in $F$, in particular at intermediate phase angles.
The asymmetry of the planet darkening events as imprinted on the change 
in polarization $\Delta P$ is due to the variation of the polarization
across the planetary disk with our model atmosphere--surface: 
the polarized flux at the limb is higher than at the terminator.

During lunar darkening events, i.e.\ planetary transits and lunar 
eclipses, the size difference between the planet and the moon 
yields a relatively symmetric change in $F$ and, due to the 
non--polarizing lunar reflection, a similarly relatively symmetric
change in $P$, as the latter is only due to the decrease in total flux,
upon a lunar darkening event, not to a change in polarized flux.  
The curves for planetary transits have steeper slopes during the 
ingress and egress phases than the curves for the lunar eclipses, because
with the latter, the moon travels through the penumbral shadow of the
planet.
The change in $P$ depends on the size and albedo of the moon, and
on the polarization signal of the planet, which itself depends on
the atmosphere-surface model and the phase angle. 
Our simulations have been performed at 
450~nm, i.e. in the blue, where the 
scattering by the gas in the Earth--like planetary atmosphere 
strongly contributes to the planet's polarization signals. 
In particular at intermediate phase angles, the polarization
signal of a gaseous atmosphere is strong, and the change in polarization 
during the lunar darkening events can reach a few percent. Indeed, 
$\Delta P \approx 1.25\% - 2.66 \%$ for 
$54\degr < \alpha_{\rm p} < 108\degr$ during planetary transits.
Note that at these phase angles, the angular distance between the planet--moon
system and the parent star is relatively large, so these angles are favorable for
the detection of reflected starlight.
For a planet with clouds in its atmosphere, the 
continuum flux phase function will have a similar shape as that for
our cloud--free planet, except the total amount of reflected flux 
will be larger (of course not at wavelengths where atmospheric gases absorb
the light). The polarization curve of a cloudy planet will show
angular features due to the scattering of light by the cloud particles,
such as the rainbow for liquid water droplets 
\citep[see e.g.][and references therein]{Karalidi12,2007AsBio...7..320B}.

The duration of a transit event depends on the orbital parameters, 
on the sizes of the planet and moon, and on the phase angle (the latter
mostly for lunar transits). In our simulations, a typical planetary 
transit takes $\sim 4$ hours both in flux and polarization. 
A lunar transit at an intermediate phase angle of 90$^\circ$, 
takes about 2~hours in flux. In polarization, the change in $P$
is apparent during a shorter period than the change in flux $F$, due to
the distribution of polarized flux across the planetary disk.
The duration of eclipse events is somewhat longer than that of 
transit events due to the diverging shape of the shadow cone. 
In our simulations, eclipse events can take up to 6~hours, where the
polarization change in the planetary eclipses is only apparent
during part of the time of the flux change.

The results presented in this paper correspond to half of the planetary 
orbit around the star. The results for the other half of the orbit will be 
similar, except that the curves will be mirrored with respect to the central
event time, because transit and eclipse ingresses and egresses will happen
over the other side of the darkened body. 

Our results show that measuring the temporal variation in $F$ and/or $P$
during transits and eclipses could provide extra information on the 
properties of a planet and/or moon and their orbits. Extracting such 
information, however, requires not only detecting such events but also
measuring the shape of the variations in $F$ and/or $P$. 
For the interpretation of such measurements, numerical simulations 
to map in more detail the influence of the physical characteristics of 
the moon and the planet (radius, albedo, atmosphere-surface
properties) and their orbital characteristics (inclination angles, ellipticity)
on the temporal variation in $F$ and $P$ are required. 
Such simulations will be targeted in future research.


%
%

\begin{appendix} 

\section{Local illumination and viewing angles}
\label{sec:computing-angles}

In this appendix, we describe the computation of the angles required to compute
the starlight that is reflected by each pixel on the planet 
(cf.\ Eq.~\ref{eq:refStokes}): 
the phase angle $\alpha$, the local viewing zenith angle $\theta_i$, 
the local illumination zenith angle $\theta_{0i}$, 
the local azimuthal difference angle $\phi_i - \phi_{0i}$, 
and the local rotation angle $\beta_i$. 
To add the computed Stokes vectors of the moon to those of the planet, 
we usually also need rotation angle $\psi$ that redefines the lunar Stokes 
vector from the lunar scattering plane to the planetary scattering plane 
(cf.\ Eq.~\ref{eq:summation}). 

\begin{figure}[b]
\centering
\includegraphics[width=0.45\textwidth]{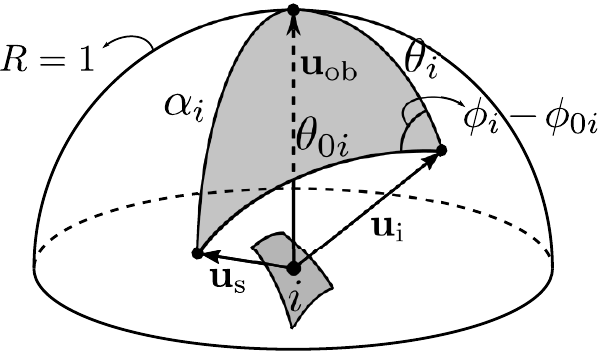}
\caption{Angular geometry of the spherical triangle centered at pixel $i$ and 
         defined by the zenith direction unit vector, $\mathbf{u}_{\rm zi}$, the 
         observer's direction unit vector, $\mathbf{u}_{\rm ob}$, and the star 
         direction unit vector, $\mathbf{u}_{\rm s}$. The sides of the spherical 
         triangle are: the observer--zenith angle $\theta_i$ , the star--zenith 
         angle $\theta_{0i}$, and the pixel--based phase angle $\alpha_i$, 
         all centered at pixel $i$. The angle between sides $\theta_{0i}$ and 
         $\theta_i$ is the azimuthal difference angle $\phi_i-\phi_{0i}$.}
\label{fig:thetatheta0}
\end{figure}

\subsection{Phase angle $\alpha_{\rm x}$}

Phase angle $\alpha$ is the angle between the direction to the star 
and the observer, as measured from the center of a body
(see Fig.~\ref{fig:thetatheta0}). In principle its value ranges from 
0$^\circ$ to $180\degr$, although the phase angle range accessible to an observer
depends on the inclination angle of the orbit of a body. A body in an edge--on
orbital geometry (inclination angle $i= 90^\circ$) can attain phase angles between
0$^\circ$ (when it is located behind the star) 
and 180$^\circ$, while a body in a face--on orbital geometry 
($i=0^\circ$) can only be observed at $\alpha=90^\circ$. Generally, given
an orbital inclination angle $i$, the phase angle range is given by
\begin{equation}
  90^\circ - i \leq \alpha \leq 90^\circ + i.
\end{equation}   
The phase angle of the planet or the moon at time $t$ is computed as
\begin{equation}
    \alpha_{\rm x}(t) = \arccos\left[ \mathbf{u}^T_{\rm z}  \cdot 
                  \left( - \frac{ \mathbf{r}_{\rm xs}(t)}
                  {\norm{\mathbf{r}_{\rm xs}(t)}} \right) \right]\,,
\label{eq:alphap}
\end{equation}
where subscript 'x' refers to either 'p' (planet) or 'm' (moon),
$\mathbf{u}^T_{\rm z}= \left[ 0, 0, 1 \right]$ 
is the unit vector along the $z$--axis, pointing towards the observer,
and vector ${\bf r}_{\rm xs}$ connects the center of the planet or moon
with the center of the star.
Given the small separation between the planet and moon compared to their 
distances to the star, $\alpha_{\rm p}$ is virtually the same as 
$\alpha_{\rm m}$, yet our numerical model uses both values.

\subsection{Local viewing zenith angle $\theta_i$}

The local viewing zenith angle $\theta_i$ is the angle between 
the zenith direction of pixel $i$ 
and the direction towards the observer (see Fig.~\ref{fig:thetatheta0}).
Angle $\theta_i$ takes values between $0\degr$ 
(at the sub-observer location) and $90\degr$ (at the limb).
It depends on the location of the pixel on the disk of the planet
or moon and is thus time-independent. It is computed according to
\begin{equation}
    \theta_i = \arccos\left[ \mathbf{u}^T_{\rm z} \cdot 
    \frac{ \mathbf{r}_{i{\rm x}}}{\norm{\mathbf{r}_{i{\rm x}}}}
    \right]\,,
\label{eq:theta_i}
\end{equation}
where subscript 'x' refers to either 'p' or 'm', $\mathbf{u}_{\rm z}$ 
is the unit vector along the $z$--axis that points towards the observer, 
and $\mathbf{r}_{i{\rm x}}$ is the vector pointing to the center of the 
pixel from the center of either the planet or the moon.
 
\subsection{Local illumination zenith angle $\theta_{0i}$}

The local illumination zenith angle $\theta_{0i}$ is defined as the 
angle between the local zenith direction of pixel $i$  
and the direction towards the star (see Fig.~\ref{fig:thetatheta0}).
Angle $\theta_{0i}$ takes values between $0^\circ$ (at the sub-stellar 
location) and $90^\circ$ (at the terminator).
The position of the star changes in time, so that the time--dependent 
local illumination zenith angle can be computed as 
\begin{equation}
    \theta_{0i}(t) = \arccos\left[ \, 
                      \frac{\mathbf{r}^T_{i{\rm x}}(t)}
                          {\norm{\mathbf{r}_{i{\rm x}}(t)}}
                     \cdot 
                     \left( -\frac{\mathbf{r}_{i{\rm s}}(t)}
                          {\norm{\mathbf{r}_{i{\rm s}}(t)}} \right)
                          \right]\,,
\label{eq:theta_0i}
\end{equation}
where subscript 'x' refers to either 'p' or 'm',  
$\mathbf{r}_{i{\rm x}}$ is the vector from the center of the 
planet or moon to the center of the pixel, and 
$\mathbf{r}_{i{\rm s}}$ the vector from the center of the star to 
the center of the pixel on the planet or moon.

\subsection{Local azimuthal difference angle $\phi_i-\phi_{0i}$}

The azimuthal difference angle $\phi_i-\phi_{0i}$ for pixel $i$ on the
planet or moon is the 
angle between the plane described by the local zenith direction and the 
direction towards the observer and the plane 
described by the local zenith direction and the direction towards the 
star\footnote[6]{Only the difference between $\phi_i$ 
and $\phi_{0i}$ is important, as our pixels are horizontally homogeneous}.
As Fig.~\ref{fig:thetatheta0} shows, $\phi_i - \phi_{0i}$ follows 
from 
\begin{equation*}
     \phi_i-\phi_{0i}(t) = \arccos{ \left( 
                    \frac{\cos{\alpha_i(t)}-\cos{\theta_i}\cos{\theta_{0i}(t)}}
                   {\sin{\theta_i}\cos{\theta_{0i}(t)}} \right) } \,.
\end{equation*}
where $\alpha_i$ is the angle between the direction to the observer and 
the direction to the star measured from the center of pixel $i$. 
Given that 
$\norm{\mathbf{r}_{i{\rm x}}} \ll \norm{\mathbf{r}_{\rm xs}}$ with 'x' 
referring to either 'p' or 'm', $\alpha_i$ can be approximated 
by the body's phase angle $\alpha_{\rm x}$, and thus
\begin{equation}
  \phi_i-\phi_{0i}(t) = \arccos{ \left(    
          \frac{\cos{\alpha_{\rm p}(t)}-\cos{\theta_i}\cos{\theta_{0i}(t)}}
                {\sin{\theta_i}\cos{\theta_{0i}(t)}} \right) } \,.
\label{eq:phi-phi0i}
\end{equation}

\subsection{Local rotation angle $\beta_i$}

Angle $\beta_i$ is used to rotate a locally computed vector 
${\bf F}^{\rm x}_i$ (see Eq.~\ref{eq:disk}) for pixel $i$
on the planet or the moon from the local meridian plane to the 
scattering plane of the body, which is used as the reference plane
for the disk--integrated signal of the body. 
The pixel grid across the planet is defined with respect to the 
planetary scattering plane, and $\beta_i$ is thus time independent
for the planetary pixels. 
For a pixel~$i$, $\beta_i$ is computed according to
\begin{equation}
   \beta_i = \arcsin{\dfrac{y_{i{\rm p}}}{x^2_{i{\rm p}}+y^2_{i{\rm p}}}} \,,
\label{eq:betai}
\end{equation}
where $x_{i{\rm p}}$ and $y_{i{\rm p}}$ are the coordinates of the center
of the pixel (recall that the $z$--axis points towards the observer).

For the lunar pixels, the alignment between the 
lunar scattering plane and the lunar grid, and hence angle $\beta_i$,
is time-dependent and requires to redefine the pixel coordinates 
with respect to the lunar scattering plane.
Indicating the redefined coordinates of lunar pixel $i$ with subscript $j$, 
angle $\beta_j$ is then computed as:
\begin{equation}
   \beta_j(t) = \arcsin{\frac{y_{j{\rm m}}(t)}
                       {x^2_{j{\rm m}}(t) + y^2_{j{\rm m}}(t)}}.
\label{eq:betaj}
\end{equation}

\subsection{Scattering plane rotation angle $\psi$}

Scattering plane rotation angle $\psi$ is used to rotate a Stokes
vector that is defined with respect to the lunar scattering plane to
the planetary scattering plane, which we use as the reference plane for the
planet--moon system.  
Angle $\psi$ is measured in the clock--wise direction from the lunar 
scattering plane to the planetary scattering plane.
For the results presented in this paper, the moon and the
planet orbit in the same, edge--on plane, and angle $\psi$ equals zero. 
In the general case, however, it is computed using
\begin{equation}
  \psi(t) = \arctan \left( -(\mathbf{u}^T_{\rm y} \cdot 
                             \mathbf{r}_{\rm ms}(t)) / 
                              (\mathbf{u}^T_{\rm x} \cdot 
                              \mathbf{r}_{\rm ms}(t)) \,  \right)\,,
\label{eq:psi}
\end{equation}
where $\mathbf{r}_{\rm ms}$ is the vector from the star to the center of the 
moon, and $\mathbf{u}_{\rm x}$ and $\mathbf{u}_{\rm y}$ are the unit vectors 
along the $x$-axis and $y$-axis in coordinate system $S_1$.

\section{Computing eclipses}
\label{sec:eclipses}

An eclipse occurs when body~$A$ is between the star $S$ and body~$B$ 
such that the shadow of~$A$ falls onto~$B$. 
The effect of a planetary or lunar eclipse depends on the positions of the star, 
moon and planet, and, due to the extended size of the star, the size, 
shape, and depth of the shadow depend not only with the radii of the star and 
the eclipsing body, but also on the distances and angles involved. 
Computing eclipses has been discussed in great detail by \citet{link69} 
for the Moon--Earth system, which we apply to our exoplanetary system. 
We model the umbral, antumbral, and penumbral shadow regions.
Figure~\ref{fig:eclipse} shows the geometries involved in the 
various types of eclipses.
The equations used for computing the influence of eclipses are 
described here.

The flux arriving at a pixel $i$ of eclipsed body~$B$ at time $t$ depends 
on the fraction of the stellar disk and the local stellar surface brightness,
as seen from the center of the pixel. In Eq.~\ref{eq:disk}, 
this is accounted for by factor $c_i$, the ratio between the actual 
flux $e'_i$ on pixel $i$ and $e_i$, the flux on the non--eclipsed pixel: 
\begin{equation}       
\label{eq:c_i}
 	c_i(t) = e'_i(t) / e_i(t) = S'_{Si}(t) / S_{Si}(t),
\end{equation}
with $S'_{Si}$ and $S_{Si}$ the stellar disk area as observed from 
pixel $i$ when it is eclipsed and when it is non--eclipsed, respectively.
Note that we ignore stellar limb darkening and stellar light that travels 
through the atmosphere of the eclipsing body (if present).

To determine $c_i$, we first have to identify whether or not 
pixel $i$ is eclipsed. Obviously, $c_i=1$ for a non--eclipsed pixel. 
If the pixel is (partly eclipsed), we have to determine the type of eclipse: 
umbral (i.e.\ total), antumbral, or annular. 
\footnote{A so--called hybrid eclipse
is an eclipse epoch where different types of eclipses occur along different
parts of the path of the shadow across the eclipsed body. Hybrid eclipses
are covered with our algorithms}.
For an umbral eclipse $c_i = 0.0$, for an antumbral and
annular eclipse, we have to compute $S'_{Si}$ in order to determine $c_i$. 

\begin{figure}[h!]
\centering 
\includegraphics[width=0.85\textwidth]{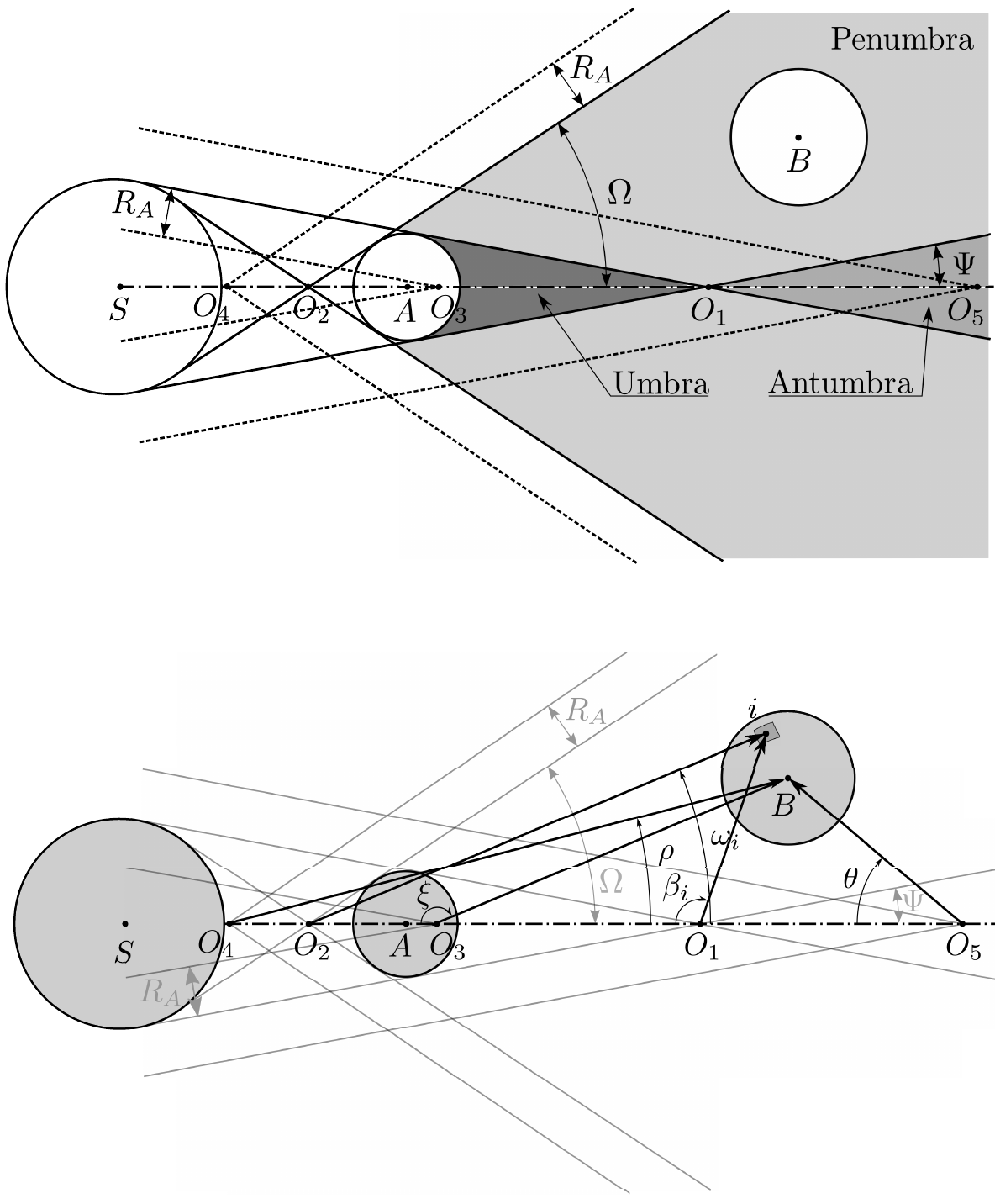}
\caption{Geometry of the umbral, antumbral and penumbral shadow cones, 
         when star $S$ is eclipsed by body $A$, casting a shadow on body $B$.
         The shadow casted by $A$ into space is rotationally symmetric 
         around the axis through the center of the star and body $A$.
         The radii of the star, body $A$ and body $B$ are denoted by 
         $R_S$, $R_A$, and $R_B$, respectively.
         Points $O_1$, $O_2$, $O_3$, $O_4$ and $O_5$ denote auxiliary 
         points: the umbral and antumbral cones have apex $O_1$ and 
         aperture $2\Psi$ and the penumbral cone has apex $O_2$ and 
         aperture $2\Omega$. The lower figure also shows angles 
         $\zeta$, $\rho$, $\omega_i$, $\beta_i$, and $\theta$, that 
         are used in the computation of the eclipse shadow depth.
         Distances between bodies and radii are not to scale in order to 
         emphasize the geometry of the system.} 
\label{fig:eclipse}
\end{figure}  

As can be seen in Fig.~\ref{fig:eclipse}, a pixel on body $B$ is eclipsed 
when it falls within the penumbral cone of body $A$. Opening angle $\Omega$ 
of the penumbral cone is given by:
\begin{equation}
    \sin{\Omega} = \dfrac{R_S+R_A}{\norm{\mathbf{r}_{\rm AS}}} \,.
\label{eq:Omega}
\end{equation}

Indeed, pixels in eclipse on the disk of body $B$ can be found at times when
\begin{equation}
  \sin{\rho}(t) < \sin{\Omega}(t),
\label{eq:eclipsecondition}
\end{equation}
with angle $\rho$ ($\left[0\degr,90\degr\right)$) given by (see Fig.~\ref{fig:eclipse}) 
\begin{equation}
 \cos{\rho} = \dfrac{-\left(\mathbf{r}_{\rm AB}-\mathbf{AO_4}\right)
              \cdot \mathbf{u}_{\rm AS}}{\norm{\mathbf{r}_{\rm AB}-\mathbf{AO_4}}}\,.
\end{equation}

Vector $\mathbf{AO_4}$ is a function of the radii of the shadowed and 
eclipsing bodies and of $\Omega$, as follows:
\begin{equation}
   \mathbf{AO_4} = \dfrac{R_B+R_A}{\sin{\Omega}}\,\mathbf{u}_{\rm AS} \,.
\end{equation}

Except when body $B$ falls completely in the penumbral cone, there will 
also be non--eclipsed pixels on the disk.
When Eq.~\ref{eq:eclipsecondition} holds, a
pixel-by-pixel search is performed, in which the center of each pixel is
checked for total or umbral (Sect.~\ref{sec:mutual-umbra}), 
annular (Sect.~\ref{sec:mutual-antumbra}), or penumbral  
(Sect.~\ref{sec:mutual-penumbra}) eclipse conditions.

\subsection{Total or umbral eclipses}
\label{sec:mutual-umbra}


In the umbral zone (see Fig.~\ref{fig:eclipse}), pixels experience a 
total stellar eclipse. If the umbral zonal is wide and the shadowed body 
$B$ relatively small, such as in the Earth--moon system, 
all pixels on the disk of $B$ can be simultaneously in the umbra,
and factor $c_i = 0$ for all pixels.
This is the case when
\begin{equation}
    \cos{\xi} > \cos{\Psi} \,,
\label{eq:conditionallumbra}
\end{equation}
with  
\begin{equation}
   \sin{\Psi} = \dfrac{R_S-R_A}{\norm{ \mathbf{r}_{\rm AS}}} \,,
\label{eq:Psi}
\end{equation}
and 
\begin{equation}
    \cos{\xi} = \frac
     {\left(\mathbf{r}_{\rm AB}-\mathbf{AO_3}\right)\cdot \mathbf{u}_{\rm AS}}
     {\norm{\mathbf{r}_{\rm AB}-\mathbf{AO_3}}} \,
     \hspace*{0.4cm} {\rm with} \hspace*{0.4cm}
     \mathbf{AO_3} = -\dfrac{R_A-R_B}{\sin{\Psi}}\,\mathbf{u}_{\rm AS}\,.
\label{eq:xi_B}
\end{equation}

The disk of body $B$ will only be partially inside the umbral shadow cone 
of body $A$ when 
\begin{equation}
    \cos{\Psi} < \dfrac{\left(\mathbf{r}_{\rm AB}-\mathbf{AO_5}\right)\cdot 
     \mathbf{u}_{\rm AS}}{\norm{\mathbf{r}_{\rm AB}-\mathbf{AO_5}}}
\end{equation}
and
\begin{equation}
    \left( \norm{ \mathbf{O_5B}} > \dfrac{R_B}{\tan{\Psi}}  
    \hspace*{0.3cm} \text{or} \hspace*{0.3cm} \norm{\mathbf{O_1B}} < R_B \right)
\label{eq:conditionumbra}
\end{equation}
Here
\begin{equation}
   \mathbf{O_5B} = \mathbf{r}_{\rm AB}-\mathbf{AO_5} \hspace*{0.5cm}
   {\rm and} \hspace*{0.5cm} \mathbf{O_1B} = \mathbf{r}_{\rm AB}-\mathbf{AO_1} \,,
\end{equation}	
with
\begin{equation}
  \mathbf{AO_5} = -\dfrac{R_B+R_A}{\sin{\Psi}}\,\mathbf{u}_{\rm AS} \hspace*{0.5cm}
  {\rm and} \hspace*{0.5cm}
  \mathbf{AO_1} = -\dfrac{R_A}{\sin{\Psi}}\,\mathbf{u}_{\rm AS} \,.
\label{eq:AO1}
\end{equation}	

If the disk of body $B$ falls partially inside the umbral cone, the pixels
where $\beta_i < \Psi$ are inside the umbra, and $c_i=0$. 
Because $c_i$ only applies to pixels on the illuminated part of the 
disk of $B$, this condition can be reformulated as
\begin{equation}
  \sin{\beta_i} < \sin{\Psi}\,.
\label{eq:betaicondition}
\end{equation}

Note that angle $\beta_i$ of a pixel can be derived from 
\begin{equation}
\cos{\beta_i} = \dfrac{\left(\mathbf{r}_{\rm Ai}-\mathbf{AO_1}\right)\cdot 
           \mathbf{u}_{\rm AS}}{\norm{\mathbf{r}_{\rm Ai}-\mathbf{AO_1}}}\,.
\end{equation}

The pixels on the disk of $B$ that are not in the umbral cone
can be in the antumbral or annular eclipse zone, as described below.

\subsection{Annular or antumbral eclipses}
\label{sec:mutual-antumbra}

Instead of crossing the umbral cone, body $B$ can cross the antumbral cone,
where eclipsing body $A$ does not completely cover the stellar disk as seen
from body $B$, thus yielding a so--called annular eclipse.
Body $B$ is in the antumbral shadow cone when 
\begin{equation}
    \cos{\xi} < -\cos{\Psi} \,,
\label{eq:xi_B_ant}
\end{equation}
where $\xi$ and $\Psi$ follow from Eqs.~\ref{eq:xi_B} and~\ref{eq:Psi}.
When Eq.~\ref{eq:xi_B_ant} is satisfied, pixels on the disk of $B$ 
are checked for their eclipsed status.
A pixel is (partially) eclipsed if 
\begin{equation}
    \cos{\beta_i} < -\cos{\Psi}\,.
\label{eq:conditionantumbra}
\end{equation}

For each pixel in the antumbral cone, factor $c_i$ is given by the fraction
of the stellar disk that is visible (see Fig.~\ref{fig:antumbra_area}), i.e.\
\begin{equation}
   c_i = \dfrac{\pi\alpha^2_{Si}-\pi\alpha^2_{Ai}}{\pi\alpha^2_{Si}} 
       = 1-\left(\dfrac{\alpha_{Ai}}{\alpha_{Si}}\right)^2\,,
\label{eq:ciantumbra}
\end{equation}
where
\begin{equation}
   \alpha_{Si} = \arcsin{\dfrac{R_{Si}}{\norm{\mathbf{r}_{Si}}}}\,
   \hspace*{0.5cm}{\rm and} \hspace*{0.5cm}
   \alpha_{Ai} = \arcsin{\dfrac{R_{Ai}}{\norm{\mathbf{r}_{Si}}}}\,.
\end{equation}

Here, $\mathbf{r}_{Si}$ is the position vector of pixel $i$ on body 
$B$ with respect to the center of star $S$.

\begin{figure}[t]
\centering
\includegraphics[width=3.5cm]{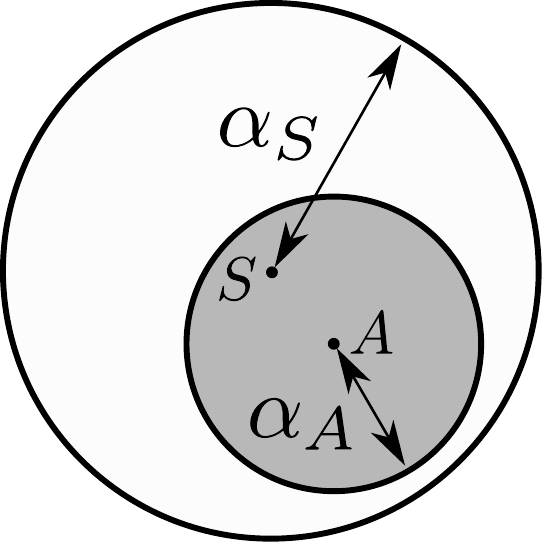}
\caption{The disks of star $S$ and eclipsing body $A$ seen 
         from a pixel on body $B$ in the antumbral zone. Angles 
         $\alpha_{S}$ and $\alpha_{A}$ indicate the angular 
         radii of the bodies.}
\label{fig:antumbra_area}
\end{figure}

\subsection{Penumbral eclipses}
\label{sec:mutual-penumbra}

When Eq.~\ref{eq:eclipsecondition} holds, all pixels that are not in the 
umbral or antumbral eclipse are examined for being in the penumbral shadow. 
Indeed, pixels with $\omega_i<\Omega$ are within the penumbral cone, as 
can be seen in Fig.~\ref{fig:eclipse}. This inequality can be 
rewritten as
\begin{equation}
   \cos{\omega_i} > \cos{\Psi}\,,
\label{eq:conditionpenumbra}
\end{equation}
where $\cos{\Psi}$ follows from Eq.~\ref{eq:Psi} and 
$\cos\omega_i$ is given by:
\begin{equation}
 \cos{\omega_i} = \dfrac{-\left(\mathbf{r}_{\rm Ai}-\mathbf{AO_2}\right)\cdot  
                \mathbf{u}_{\rm AS}}{\norm{\mathbf{r}_{\rm Ai}-\mathbf{AO_2}}}\,
                \hspace*{0.4cm} {\rm with} \hspace*{0.4cm}
  \mathbf{AO_2} = \dfrac{R_A}{\sin{\Omega}}\,\mathbf{u}_{\rm AS}\,.
\end{equation}

The magnitude of the eclipse 
at pixel $i$, $c_i$, can be calculated through the stellar and 
eclipsing body viewing angles, i.e. the angular diameter of the 
bodies, $2\alpha_S$ and $2\alpha_A$, and the eclipsing body-to-star 
angular distance as seen from the shadowed body, $\delta$. Then, as 
follows from Fig.~\ref{fig:area_eclipse}, $c_i$ can be computed using
\begin{equation}
  c_i = \dfrac{\pi\alpha_S^2-A_1-A_2}{\pi\alpha_S^2} = 
        1- \dfrac{A_1+A_2}{\pi\alpha_S^2},
\label{eq:cipenumbra}
\end{equation}
with
\begin{equation}
   A_1 = \left\{ {\begin{array}{*{20}{l}}
         \dfrac{\theta_A-\sin{\theta_A}}{2} \hspace*{0.1cm}
         \alpha_A^2 \hspace*{1.0cm} \text{if} \hspace*{0.2cm} \delta\ge l_S \\ \\
         \pi - \dfrac{\theta_A-\sin{\theta_A}}{2} \hspace*{0.1cm}
         \alpha_A^2 \hspace*{0.45cm} 
         \text{if} \hspace*{0.2cm} \delta< l_S
\end{array}} \right. \,
\end{equation}
and
\begin{equation}
   A_2 = \left\{ {\begin{array}{*{20}{l}}
      \dfrac{\theta_S-\sin{\theta_S}}{2} \hspace*{0.1cm}
      \alpha_S^2 \hspace*{1.0cm} \text{if} \hspace*{0.2cm} \delta\ge l_A \\ \\
      \pi - \dfrac{\theta_S-\sin{\theta_S}}{2} \hspace*{0.1cm}
      \alpha_S^2 \hspace*{0.45cm} 
      \text{if} \hspace*{0.2cm} \delta< l_A
\end{array}} \right. \,
\end{equation}

\begin{figure}[t]
\centering
\includegraphics[width=7cm]{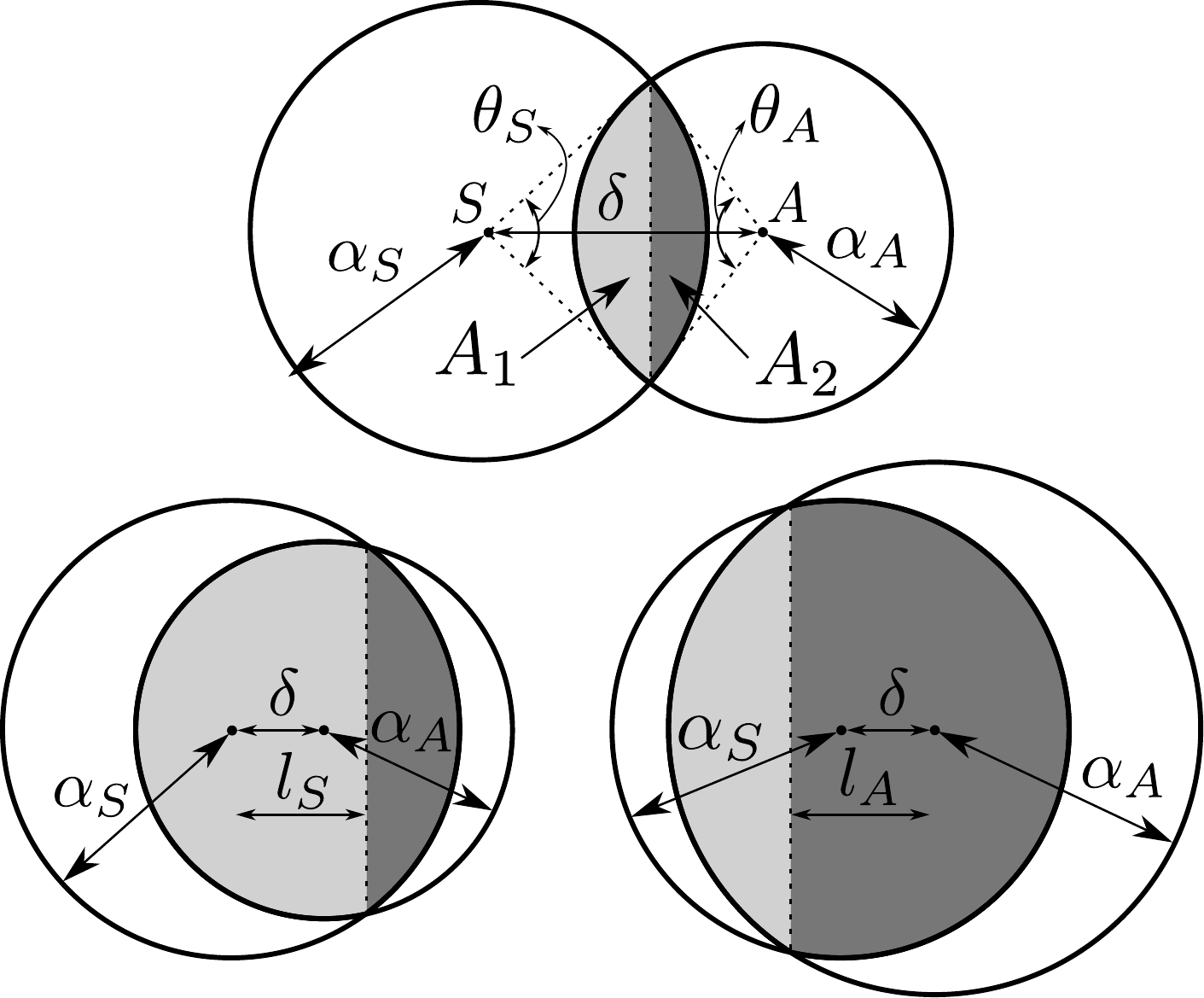}
\caption{Stellar ($S$) and eclipsing body ($A$) disks as seen from a 
         pixel on the shadowed body during a penumbral eclipse. The stellar 
         shadowed area is decomposed in two components $A_1$ and $A_2$. 
         $\alpha_S$ and $\alpha_A$ stand for the angular radius of the bodies, 
         and $\delta$ is the angular separation between the bodies' center. 
         $\theta_S$ and $\theta_A$ are the central angles of the circular 
         segments defined by the common cord of the intersecting stellar and 
         eclipsing body disks. The minimum distance from the star/eclipsing 
         body center to the common chord is defined as $l_S$/$l_A$.}
\label{fig:area_eclipse}
\end{figure}

Here, $l_S$ and $l_A$ are the distances from the centers of the stellar disk  
and eclipsing body $A$, respectively, to the line between the two points
where the stellar and eclipsing body intersect. The angles $\theta_S$ and 
$\theta_A$ (see Fig.~\ref{fig:area_eclipse}) follow from e.g.\ Heron's formula:
\begin{equation}
 \theta_S = 2\arcsin{\left( \dfrac{2}{\alpha_S\delta}
            \sqrt{S\left(S-\alpha_S\right)\left(S-\alpha_A\right)
            \left(S-\delta\right)} \right)}\,,
\end{equation}
and
\begin{equation}
 \theta_A = 2\arcsin{\left( \dfrac{2}{\alpha_A\delta}
            \sqrt{S\left(S-\alpha_S\right)\left(S-\alpha_A\right)
            \left(S-\delta\right)} \right)}\,,
\end{equation}
with
\begin{equation}
  S = \dfrac{1}{2}\left(\alpha_A+\alpha_S+\delta\right)\,.
\end{equation}

Pixels on body $B$ that satisfy Eq.~\ref{eq:eclipsecondition} but do 
not meet the conditions established by Eqs.~\ref{eq:conditionallumbra}, 
\ref{eq:conditionumbra}, \ref{eq:conditionantumbra} and 
\ref{eq:conditionpenumbra} are not eclipsed and have $c_i=1.0$.

\section{Number of pixels across the disk}
\label{app:NN}

The results of our numerical simulations depend on the number of pixels 
that is used to compute the flux and polarization 
signals of the disk of the planet and moon. 
The number of pixels across the disk of the planet or the moon,
$N^{\rm p}$ and $N^{\rm m}$, respectively, determines the spatial resolution 
of the locally reflected Stokes vectors and hence the numerical error in the
integration across the disk, in particular at large phase angles. 
However, the larger the number of pixels, the smaller the error but the longer 
the computational time.
We have investigated the optimal number of pixels, expressed in the number 
of pixels across the equator of the planet and the moon, $N_{\rm eq}^{\rm p}$ 
and $N_{\rm eq}^{\rm m}$, respectively, using a trade--off between the errors and 
the computational time, with time steps of 24 hours. 

For the trade--off, we compute the flux $F$ (Eq.~\ref{eq_APP_Fa}) 
and degree of polarization $P$ (Eq.~\ref{eq_APP_Pa}) for consecutive 
values of $N_{\rm eq}$ as functions of phase angle $\alpha$, both for the
planet and the moon. In Fig.~\ref{fig:appendix_Neq}, we show the maximum
and mean differences encountered across the whole phase angle range of the 
planet and the moon, together with the average 
disk integration time and the total phase curve computation time.
The differences are defined as 
\begin{eqnarray}       
    |\Delta F_{n}(\alpha)| & = & 
    \dfrac{|F_{n}(\alpha)-F_{n-1}(\alpha)|}{F_{n-1}(\alpha=0\degr)}, \\
\label{eq_APP_Fa}       
    |\Delta P_{n}(\alpha)| & = & |P_{n}(\alpha)-P_{n-1}(\alpha)|,
\label{eq_APP_Pa}    
\end{eqnarray}
with $n-1$ and $n$ two consecutive values of $N_{\rm eq}$.

Figure~\ref{fig:appendix_Neq} shows that the flux and polarization
differences decrease with increasing $N_{\rm eq}$, and that for the 
values of $N_{\rm eq}$ considered, the computed flux and polarization 
curves have not yet completely converged. However, further increasing 
$N_{\rm eq}$ increases the integration time across the planetary disk, as 
can be seen in Fig.~\ref{fig:appendix_Neq}c. 
In the simulations presented in this paper, we decided to use 
$N_{\rm eq}^{\rm p}= 50$ and $N_{\rm eq}^{\rm m}= 14$, which yields an 
average disk--integration time of $\sim 0.8$ seconds (thus an overall phase curve 
computation time of $\sim 2.4$ minutes with a 24~h temporal resolution).
These values for $N_{\rm eq}^{\rm p}$ and $N_{\rm eq}^{\rm m}$ produce 
smooth curves for individual transit and eclipse events for temporal 
resolutions as small as 1~minute. 

\begin{figure}[h]
\centering
\includegraphics[width=\textwidth]{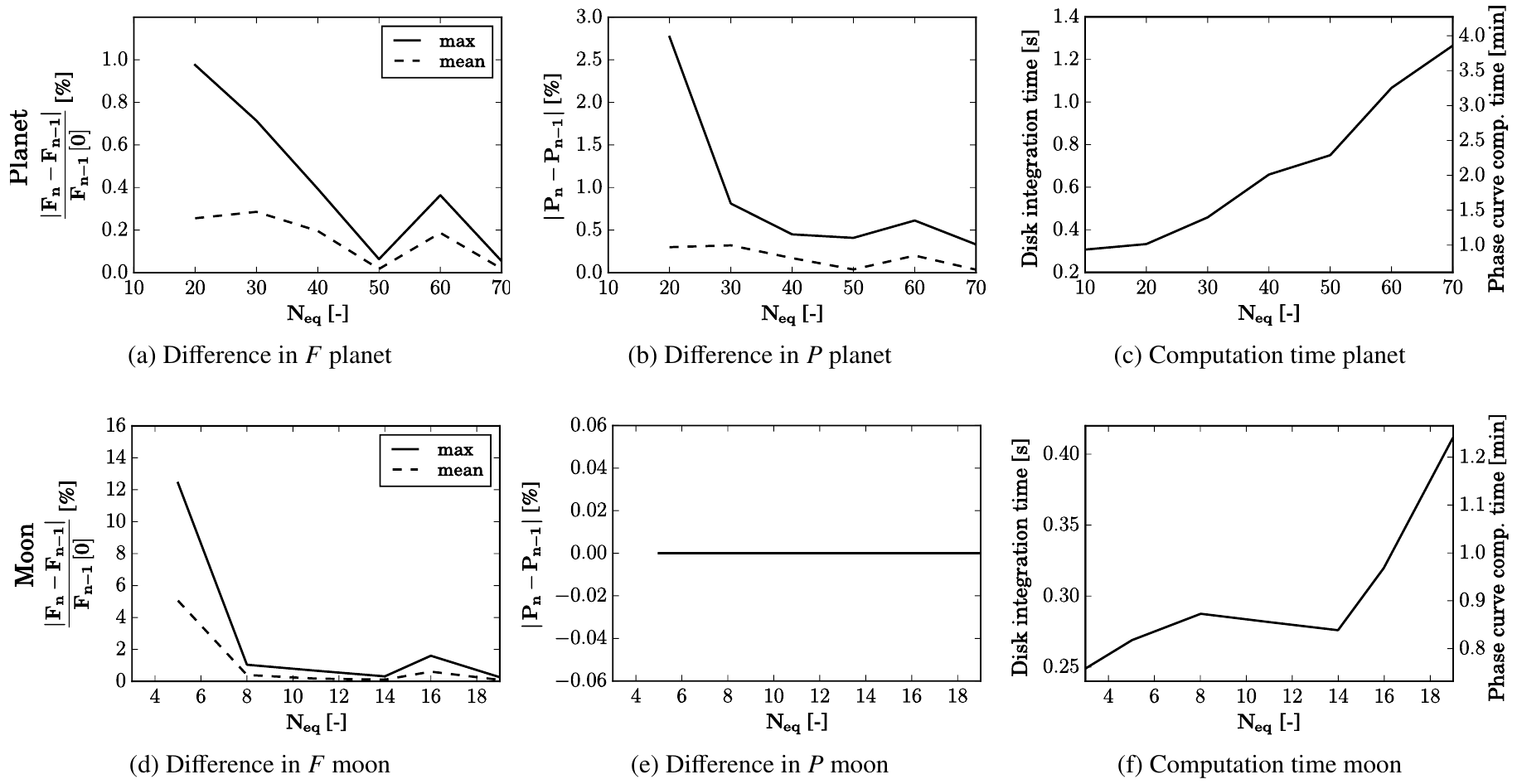}
\caption{Analysis for the number of pixels along the equator 
	     $N_{\rm eq}$ of the planet (top) and moon (bottom). 
         Shown are the maximum (solid line) and mean (dashed line) differences
         between results computed across the whole phase angle range
         and for consecutive values of $N_{\rm eq}$ values, for the reflected
         flux $F(\alpha)$ relative to $F(\alpha=0^\circ)$ (a and d), 
         and degree of polarization $P$ (b and e, note
         that for the moon $P= 0$). Also shown is the computational time
         (in minutes) for the computation of a full phase curve (with 
         24 hour temporal resolution) and the
         average disk integration (c and f). For $N_{\rm eq}^{\rm p}$
         (top), we used values of 10, 20, 30, 40, 50, 60, and 70, and 
         for $N_{\rm eq}^{\rm m}$ (bottom), 3, 5, 8, 11, 14, 16, and 19.
         } 
\label{fig:appendix_Neq}
\end{figure}
\end{appendix}

\end{document}